\documentclass{aa}  
\usepackage{graphicx}
\usepackage{txfonts}

\begin{document}

   \title{Properties of the diffuse gas component in filaments detected in the Dianoga cosmological simulations}

   \author{Samo Ilc
          \inst{1}
          \and
          Dunja Fabjan
          \inst{1,2,3}
          \and
          Elena Rasia
          \inst{2,3}
          \and
          Stefano Borgani
          \inst{4,2,3,6,5}
          \and
          Klaus Dolag
          \inst{7,8}
          }

\institute{
Faculty of Mathematics and Physics, University of Ljubljana, Jadranska cesta 19, 1000 Ljubljana, Slovenia
\and
INAF – Osservatorio Astronomico di Trieste, via G. B. Tiepolo 11, I-34143 Trieste, Italy
\and
IFPU – Institute for Fundamental Physics of the Universe, Via Beirut 2, I-34014 Trieste, Italy
\and
Department of Physics, Astronomy Section, University of Trieste, via G. B. Tiepolo 11, I-34131 Trieste, Italy
\and
INFN, Instituto Nazionale di Fisica Nucleare, Via Valerio 2, I-34127, Trieste, Italy
\and
ICSC - Italian Research Center on High Performance Computing, Big Data and Quantum Computing, via Magnanelli 2, 40033, Casalecchio di Reno, Italy
\and
Universitäts-Sternwarte, Fakultät für Physik, Ludwig-Maximilians-Universität München, Scheinerstr.1, 81679 München, Germany 
\and 
Max-Planck-Institut für Astrophysik, Karl-Schwarzschild-Straße 1, 85741 Garching, Germany
}

   \date{Received  March 23, 2024; accepted July 27, 2024}

 
  \abstract
        {Cosmic filaments are observationally hard to detect. Hydrodynamical cosmological simulations are ideal laboratories where the evolution of the cosmic web can be studied. This allows for easier insight into the nature of the filaments.}
        {We investigate how the intrinsic properties of filaments are evolving in areas extracted from a larger cosmological simulation. We aim to identify significant trends in the properties of Warm-Hot Intergalactic Medium (WHIM) and suggest possible explanations.}
        {To study the filaments and their contents, we select a subset of regions from the Dianoga simulation. We analysed these regions that were simulated with different baryon physics, namely with and without the AGN feedback. We construct the cosmic web using the Sub-space Constrained Mean Shift (SCMS) algorithm and the Sequential Chain Algorithm for Resolving Filaments (SCARF). We examined the basic physical properties of filaments (length, shape, mass, radius) and analysed different gas phases (hot, WHIM and colder gas components) within those structures. The evolution of the global filament properties and the properties of the gas phases were studied in the redshift range $0 < z < 1.48$.}
        {Within our simulations, the detected filaments have, on average, lengths below $9$ Mpc. The filaments' shape correlates with their length; the longer they are, the more likely they are curved. We find that the scaling relation between mass $M$ and length $L$ of the filaments is well described by the power law $M \propto L^{1.7}$. The radial density profile is widening with redshift, meaning that the radius of the filaments is getting larger over time. The fraction of gas mass in the WHIM phase does not depend on the model and is rising towards lower redshifts. However, the included baryon physics has a strong impact on the metallicity of gas in filaments, indicating that the AGN feedback impacts the metal content already at redshifts of $z \sim 2$.}
   {}


   \maketitle

\section{Introduction}
\label{sec:Intro}
    
    Clusters of galaxies lie at the intersection of the filamentary structure that forms the cosmic web \citep{Bond96}. While cluster properties have also been intensively studied in a broader cosmological context, the filamentary structure, i.e., both the large-scale cosmic filaments and bridges connecting pairs of clusters, is still poorly understood. The complex evolution of the cosmic web can be studied by describing the properties of the large-scale filaments and shorter bridges between galaxy clusters.

    In the last few decades, the search for large-scale filaments has taken a big leap forward, with a compilation of a number of filament catalogues (usually based on the galaxy distribution), e.g. within the Sloan Digital Sky Survey (SDSS) \citep{Carron22}, the two degree Field Galaxy Redshift Survey (2dF GRS) \citep{Pimbblet2004}, the Galaxy and Mass Assembly survey (GAMA) \citep{Alpaslan14a}, in the Cosmic Evolution Survey (COSMOS) field \citep{Luber2019}, the Sydney-AAO Multi-Object IFS (SAMI) Galaxy Survey \citep{Welker2020}, the VIMOS Public Extragalactic Redshift Survey (VIPERS) \citep{Malavasi2017}.

    As for the gas component tracing the cosmic filaments, it is currently difficult to detect due to the low density and low emissivity that diffuse baryons have within filaments. The gas component in the filamentary structures, such as intercluster filaments and larger cosmic filaments, is studied by combining different data and using different techniques. For example, ROSAT X-ray data \citep{Tanimura20nov} and stacked Compton-y maps from Planck satellite \citep{Tanimura20may} were used to obtain gas temperatures and overdensities for a large number of cosmic filaments at intermediate redshifts (detected previously using SDSS data) having lengths above $30$ Mpc. A study of the network of filaments around the Coma cluster was performed by \cite{Malavasi2020} combining SDSS data with the cosmic web detection algorithm DisPerSe \citep{Sousbie11}, obtaining information about cluster connectivity and a tentative detection of Sunyaev-Zeldovich (SZ) signal within filaments using Planck data. To analyse the signal of hot gas in extended structures \cite{Lokken2022} combined Compton-y maps from the Atacama Cosmology Telescope (ACT) stacked on redMaPPer cluster positions from the optical Dark Energy Survey (DES).
    
    Gas in bridges between clusters is commonly studied in X-rays \citep[e.g.][among others]{Sugawara17, Alvarez18} combining data from Chandra, XMM-Newton, Suzaku and recently from eRosita. Some bridges (between Abell 399 and Abell 401) were also studied using thermal SZ effect and performing a multiwavelength analysis \citep[e.g.][for example]{Bonjean18, Hincks22}.
    
    Focusing on the intergalactic gas, according to numerical simulations, at present epoch $\simeq 40-50$\% of the baryons hosted within filaments are in the form of a Warm-Hot Intergalactic Medium (WHIM) \citep{CenOstriker06, Dave2001}. In the last decade, a number of observations have focused on the detection and characterization of the diffuse WHIM in the vicinity of clusters \citep[e.g.][]{Takei07, Akamatsu11, Eckert15, Bulbul2016} and between pairs of clusters or cluster systems, e.g. A222/223 \citep{Werner08}, A3391/A3395 \citep{Reiprich21, Veronica2024}, A2029/A2033 \citep{Mirakhor22}, Abell 98 triple merging system \citep{Alvarez2022}, A399/401 \citep{Akamatsu2017, Bonjean18}. Recently, \cite{Reiprich21} analyzed the complex galaxy cluster system Abell 3391/95 using SRG/eROSITA data and found that it contains (among other structures) a warm-hot emission filament $15$ Mpc long. \cite{Zhang2024} used data from the eRASS All-Sky X-ray survey combined with the SDSS optical filament catalogue, detecting WHIM by stacking around $8000$ X-ray filaments. Mild positive detections of the WHIM in cosmic filaments were reported using the X-ray stacking analysis of cosmic filaments \citep{Tanimura20nov, Tanimura2022} and stacked SZ effect signal from intergalactic medium \citep{deGraaff2019,Tanimura20may}.
 
    In simulations, the detection of filaments can be done with different methods. Usually, four main components are detected as parts of the cosmic web: knots, filaments, sheets, and voids. In the past years, a number of different methods were employed to trace the large-scale structure and its components. Some of the methods are able to detect all the objects of the cosmic web, e.g. NEXUS+ \citep{Cautun13}, T-web \citep{Forero09} and V-web \citep{Hoffman12}. Similarly, DisPerSE \citep{Sousbie11} and Spineweb \citep{Aragon10} detect all the cosmic web elements, but instead of knots, they map topological nodes. Some of the methods are able to obtain all but knots, e.g. MMF-2 \citep{Aragon14}, while others focus only on filaments, e.g. SCMS \citep{Chen15}, MST \citep{Alpaslan14a} and Bisous \citep{Tempel14}. Despite the different approaches, all of the methods give overall comparable results, with very similar structures detected. For a detailed comparison and more in-depth analysis of the main methods, see \cite{Libeskind18}.

    With a large number of available methods to detect the cosmic web, cosmological simulations are nowadays used to characterise the properties of filaments \citep[e.g.][]{Cautun14,Galarraga20,Galarraga21}, their evolution \citep[e.g.][]{Zhu21,Galarraga24}, the properties of galaxies within filaments \citep[e.g.][]{Lee21, Zakharova2023} and the connection between the cluster outskirts and filaments \citep[e.g.][]{Rost21, Kuchner21}. \cite{Angelinelli21} studied clumps in and around galaxy clusters in non-radiative simulations. Since high-density clumps are easier to detect with X-ray telescopes, they expect them to be tracers of the filaments that compose the cosmic web. They find the filament temperature to (mildly) correlate with the mass of the main cluster. Besides studies of the properties of the cosmic web, simulations are also used to investigate the origin of the gas in the observed filamentary regions \citep[e.g.][]{Biffi22} and to predict future observational strategies \citep[e.g.][for the detection of WHIM in the soft X-ray band]{Churazov23}.

    In this work, we used a sample of resimulated volumes extracted from a larger cosmological simulation to investigate the intrinsic properties of the filaments within those regions. By selecting regions with different accretion histories and comparing simulations where different feedback effects are at work, we aim to study the mechanisms that impact the thermal and chemical properties of the intergalactic gas within the detected filaments, focusing on the WHIM phase.
    
    In Sect. \ref{sec:data}, we describe the simulations used for this work. Sect. \ref{sec:detection} is devoted to the detailed description of the methods used to detect filaments with the Sub-space Constrained Mean Shift (SCMS) algorithm and how to determine their properties. In Sect. \ref{sec:Results}, we explore the physical properties of the filaments and their evolution in time. The last Sect. \ref{sec:conclusions} summarizes our conclusions.

\section{Data}
\label{sec:data}

    \subsection{The simulation code}
    \label{sub:dianoga}
    
    In this work, we focused on a subset of Lagrangian regions that are centred around massive galaxy clusters and are part of the Dianoga set of simulations. These zoomed-in regions were extracted from a parent dark matter-only cosmological simulation (of h$^{-3}$ $1$ Gpc$^3$) originally described in \cite{Bonafede11} and later re-simulated with an improved resolution with the addition of the baryonic component. The set of simulated regions adopted for this work was performed with GADGET-3, which is a modern version of the Tree-PM Gadget code \citep{Springel05} and accounts for a new description of the smooth particle hydrodynamics that includes higher order interpolation kernels (Wendland C$^4$ kernel with $200$ neighbours) and advanced formulations for artificial viscosity and thermal diffusion \citep{Beck16}. The full set of simulations was described initially in \cite{Rasia15} and extensively studied in a number of later works \citep[see, e.g.][]{Planelles17, Biffi17, Biffi18, Truong18}.
    
    \begin{table}
    	\centering
    	\caption{Main properties of the regions used in this work. }
    	\label{tab:Regions}
    	\begin{tabular}{l|ccccc}
    		\hline
    		reg.  & M$_{\rm 200}$          & R$_{\rm 200}$ & N$_{\rm cl}$ & Size           & V   \\
    		     & $10^{14}$ M$_{\odot}$  & Mpc           & & R$_{\rm 200}$        
                 &($10$ Mpc$)^3$  \\

            \hline
                &  \multicolumn{5}{c}{AGN} \\
    		\hline
            D1 & 28.71  & 4.63 & 19 & 14.64 & 311.03  \\
            D5 & 2.86   & 2.14 & 18 & 22.47 & 110.64  \\
            D6 & 24.48  & 4.39 & 30 & 16.31 & 363.47  \\
            D9 & 2.17   & 1.96 & 14 & 25.35 & 117.50  \\
            D22& 31.43  & 4.77 & 97 & 20.13 & 881.30  \\
    		\hline
                &  \multicolumn{5}{c}{CSF} \\
    		\hline
            D1 & 28.44  & 4.62 & 22  & 14.68 & 310.96  \\
            D5 & 2.79   & 2.13 & 18  & 22.64 & 110.55  \\
            D6 & 24.33  & 4.38 & 29  & 16.35 & 363.57  \\
            D9 & 2.16   & 1.95 & 15  & 25.41 & 117.60  \\
            D22& 31.17  & 4.76 & 98  & 20.19 & 881.32  \\
    		\hline
    	\end{tabular}
    \tablefoot{The table collects some of the general properties of each region: the region identifier \citep[as in][]{Roncarelli2013}, the total mass (evaluated at $R_{\rm 200}$, where this radius corresponds to $200$ times the critical density, $\rho_c$) $M_{\rm 200}$ of the main (central) cluster (in units of $10^{14} M_{\odot}$), $R_{\rm 200}$ of the main cluster in units of Mpc, number of groups and clusters $N_{\rm cl}$ that have masses $M_{\rm 200} > 10^{13} M_{\odot}$, mean side of the cubic box in units of $R_{\rm 200}$ and the approximate cubic volume $V$ centred on the main cluster in units of $(10$ Mpc$)^3$ (physical units). The properties in this table are evaluated at $z=0$ and reported for both AGN and CSF simulations.}
    \end{table}
    
    Simulations presented here are based on a $\Lambda$CDM model, with cosmological parameters consistent with 7-year WMAP measurements \citep[see][]{Komatsu11}: $\Omega_m = 0.24$ and $\Omega_b = 0.04$ for the density parameters of matter and baryons, $H_0 = h_0 \; 100$ km s$^{-1}$ Mpc$^{-1}$ and $h_0=0.72$ for the present day Hubble parameter, $n_s = 0.96$ for the primordial spectral index and $\sigma_8 = 0.8$ for the amplitude of the power spectrum of the density fluctuations. The selected Lagrangian regions are extracted and re-simulated with the zoomed-initial technique (ZIC) described in \cite{Tormen97} by increasing the resolution in mass. In the high-resolution region, gravity is calculated with a Plummer-equivalent softening length of $\varepsilon = 2$ h$^{-1}$ kpc for stars and black holes and $\varepsilon = 3.75$ h$^{-1}$ kpc for DM and gas particles. The softening is fixed to comoving coordinates for all except DM particles, to which below $z = 2$ is given in physical units. The highest mass resolution of DM particles in this set of simulations is $m_{\rm DM} = 8.47 \times 10^8$ h$^{-1}$ M$_{\odot}$, while the initial mass of a gas particle is $m_{\rm gas} = 1.53 \times 10^8$ h$^{-1}$ M$_{\odot}$. 
    
    The set of Lagrangian regions was re-simulated with two different baryon physics models. The main difference between the two is the presence (or absence) of feedback from Active Galactic Nuclei (AGN). In particular:
    \begin{itemize}
        \item CSF (Cooling and Star Formation) model - This model accounts for radiative cooling and subsequent formation of star particles. Star particles that describe the evolution of a stellar population include in the subgrid model also the feedback from supernovae (SN) and the metal enrichment from different stars (SN Ia, SN II and asymptotic giant branch AGB stars) \citep[see, e.g.][for a detailed explanation]{Biffi17,Biffi18}. Stellar evolution and metal enrichment models are described in \cite{Tornatore07} and allow to follow the distribution of fifteen chemical species (H, He, C, Ca, O, N, Ne, Mg, S, Si, Fe, Na, Al, Ar and Ni). These elements also contribute to the cooling. Models by \cite{Wiersma2009} and \cite{Haardt2021} account for rates of metal-dependent radiative cooling and for the effects of the UV/X-ray background emission. Prescriptions for star formation are based on the work by \cite{Springel2003}, where the velocity of galactic winds $v_w$ originated by SN driven outflows, which are fixed at $350$ km/s.
        \item AGN (Active Galactic Nuclei) - These simulations rely on the same prescriptions described in the CSF case, and, in addition, they include the feedback effect from AGNs. The AGN feedback is based on a subgrid model of accretion on supermassive black holes (SMBH), where the mechanical and radiative outflows are accounted for as thermal feedback in both cases. Kinetic feedback from jets would be required to resolve sub-kpc scales and is not modelled explicitly \citep[see][]{Steinborn2015}. The gas accretion rate (Eddington-limited) and the SMBH mass are both included in the efficiencies of the two outflows and provide a smooth transition between the radio and quasar mode. The model can account for hot and cold accretion, but in Dianoga simulations \citep[see][]{Rasia15}, only cold gas accretion is considered. We refer to \cite{Steinborn2015} for a detailed description of the model and its performance in general.
    \end{itemize}
    
    \subsection{The cluster sample}
    Table \ref{tab:Regions} contains the main characteristics of the high-density regions selected for the analysis. Three of the regions (namely D1, D6 and D22) contain a central cluster with mass $M_{\rm 200} > 1.7 \cdot 10^{15} M_{\odot} \;  h^{-1}$, while the main clusters in the other two regions (D5 and D9) have central clusters with $\sim 10$ times smaller masses. The latter regions were selected as "isolated regions", and their present-day volume is three times smaller than that of the largest and denser regions. We also checked the merging histories of the central clusters. In all but one case, the last major merger\footnote{Major merger is here defined as a merging event, in which the less massive progenitor is at least one-fourth of the more massive one.} happened at lookback time of about $6.5$ Gyr or greater (corresponding to $z > 0.7$). The only exception is region D6, where the last major merger happened $3.77$ Gyrs ago (at $z \simeq 0.3$).

    Inside each Lagrangian region, the code identifies the main halos using a Friend-of-Friend (FoF) algorithm, while substructures within haloes are detected with the \textsc{Subfind} algorithm \citep{Springel2001}. The identification of self-bound substructures inside FoF halos was defined in its original form for DM-only simulations and extended to simulations with baryon physics by \citet{Dolag2009}. The sample of main haloes is obtained with a FoF algorithm with a linking length equal to $0.16$ times the mean dark matter particle separation. The centre of each halo is defined by the most bound particle (the particle with the lowest gravitational potential). Radii and masses at different overdensities are then calculated around each cluster centre, while substructures are identified within the cluster virial radius. Filaments are detected within each region using the centres of substructures.

\section{Determining the filaments and their properties}
\label{sec:detection}

\subsection{The Sub-space Constrained Mean Shift (SCMS) algorithm}
\label{sec:SCMS}

    The Subspace Constrained Mean Shift (SCMS) algorithm, described by \cite{Ozertem2011}, was modified and used as a method for filament detection by \cite{Chen15}. This method models filaments as ridges of the galaxy probability density function. The same algorithm has been successfully used for addressing different problems: reconstructing real cosmic filaments using SDSS \citep{Carron22}, studying galaxy-filament alignment \citep{Chen19sdss}, filaments in the hydrodynamic simulation MassiveBlack-II\footnote{See \href{https://sites.google.com/site/yenchicr/home}{Cosmic Web Reconstruction page} for references and filament catalogues obtained in the MassiveBlack-II simulation.} \citep{Chen15sim}, weak lensing maps \citep{Moews21} and studying velocity around stellar filaments \citep{CrisPy2020}. 
    
    SCMS is a multiple-step algorithm that determines filaments based on the density function $p(x)$, where $x$ is the spatial coordinate. The density function can be calculated with the standard kernel density estimator (KDE)
    \begin{equation}
        p(x) = \frac{1}{nh^{d}} \sum_{i=1}^{n} K\bigg(\frac{||x-X_{i}||}{h}\bigg),
        \label{eq:Eq1}
    \end{equation}

    where $n$ is the number of tracers with coordinates $X_i$ and $||x-X_{i}||$ is the Euclidean distance between the $i$ tracer and the location where KDE is evaluated; $d$ is number of spatial dimensions (e.g. $d=2$ or $d=3$), $h$ is the smoothing bandwidth, $K$ is the smoothing (e.g. Gaussian) kernel. The details of the algorithm can be found in Appendix \ref{app:A} and in \cite{Chen15} while here below, we specify the input values that we chose for this work:

    \begin{itemize}
         \item Tracer coordinates: as previously said, we aim at identifying the filaments from the galaxy probability function to follow an observational-like approach; thus, our tracers are the substructures identified with \textsc{Subfind}. Since we select them by applying a cut in total mass, they can be associated with galaxies (there will be a subtle difference in the selection function). Specifically, we consider all substructures with total masses $> 10^{11}$  h$^{-1}$M$_{\odot}$, which is a good compromise between the resolution of our simulations and the need for large statistics of tracers.
 
        \item Smoothing bandwidth $h$: this parameter controls the smoothing of the distribution of tracers and thus depends on the local property of the density field. We utilize the expression provided by \cite{Chen15}:
        \begin{equation}
            h = A_0 \bigg(\frac{1}{d+2}\frac{1}{n}\bigg)^{\frac{1}{d+4}} \sigma_{min},
            \label{eq:smooth}
        \end{equation}
    
        where $n$ and $d$ have the same meaning as before, $\sigma_{min}$ is the minimal value of the three standard deviations associated with the spatial coordinate, and $A_0$ is the only parameter that does not depend on the local environment and needs to be calibrated according to the analyzed simulation. In our case, we chose $A_0=0.5$ after visually inspecting the overlap of the resulting skeleton and the tracer density field (see Appendix \ref{app:B}). The usual value of the smoothing bandwidth $h$ is in the range $1-1.6$ Mpc.

        \item Threshold parameter $\tau$: when defining a filament in a selected region, we can a priori avoid the computation in lowest-dense volumes. This translates to discarding all $x$ locations whose density function is below a certain threshold,  $p(x)<\tau$. Applying this selection avoids false filament detections, reduces overall noise, and speeds up the process. The value of the threshold parameter again depends on the local properties of the environment. Namely, it is defined as the difference between the local, $\hat{p}$, and the mean density, $\bar{p}$:
        \begin{equation}
            \tau = \sigma(\hat{p}) \equiv \bigg(\int_\mathbb{K} (\hat{p}(x) - \bar{p}(\mathbb{K}))^2 dx \bigg)^{1/2} \sim \hat{p} - \bar{p},
        \end{equation}
    
        where $\mathbb{K}$ is the region to which we are applying the algorithm, and the local density is computed in 1 cubic cell with a size equal to the size of $\mathbb{K}$ divided by 100 (see definition of $\mathcal{M}$ in Appendix \ref{app:A}).  
   
    \end{itemize}

\begin{figure}
        \centering
        \includegraphics[width=0.95\linewidth]{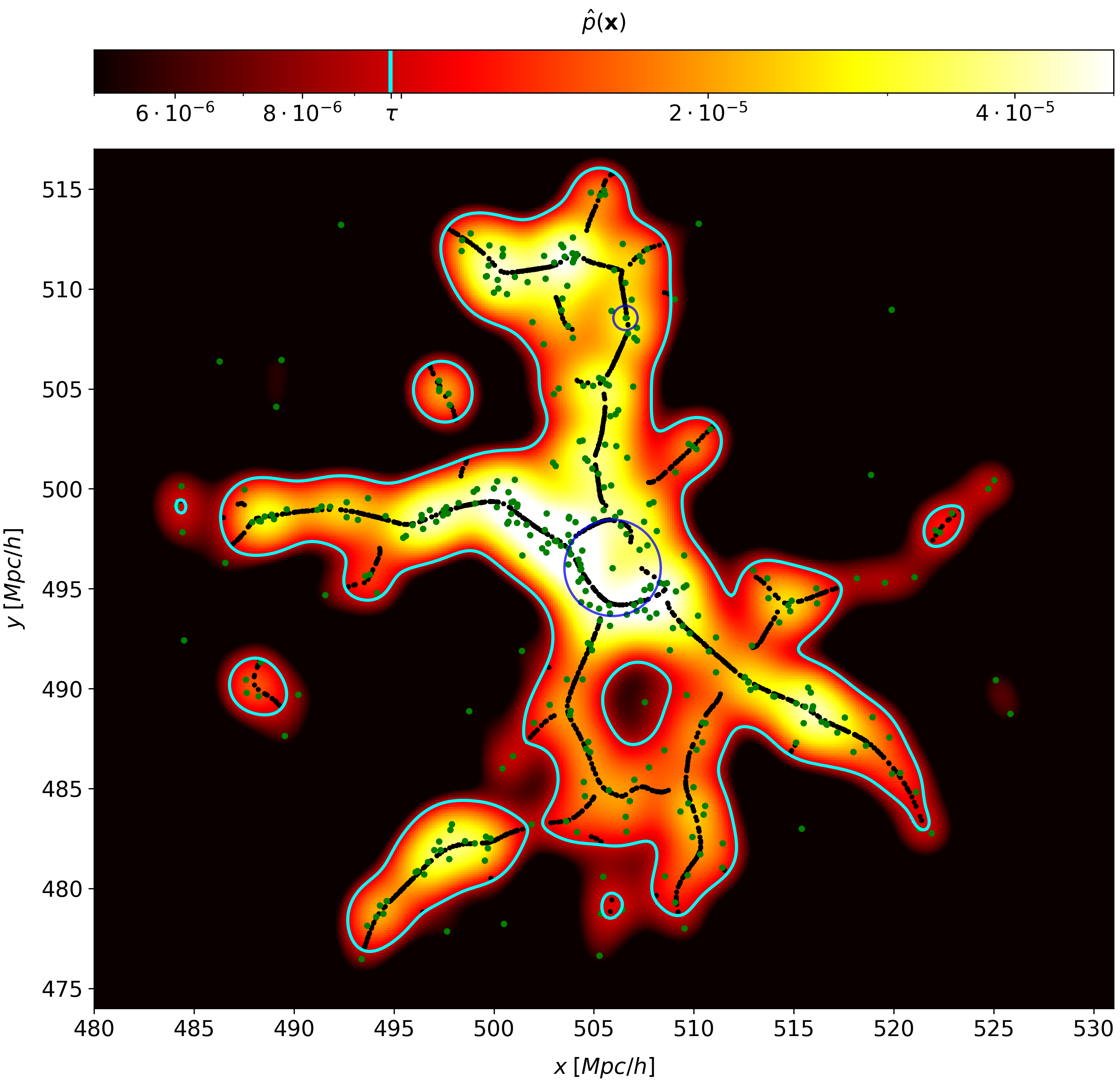}
        \caption{An example of the final result of the SCMS algorithm in 2D space. Black dots mark the points of the skeleton. Along with the skeleton, the colour-coded kernel density estimator is plotted. Tracers are marked with green points. The contour line is plotted with cyan colour at the threshold $\tau$ value. The blue circles represent the virial radius of the two massive clusters (M$_{\rm vir} > 10^{13} M_\odot$ in the 2D space. This skeleton was extracted from a thin slice ($\sim 1.2$ Mpc thick) in region D6 at redshift $z=0$.}
        \label{fig:SCMS}
    \end{figure}

    The SCMS algorithm works on a mesh of points that are shifted towards the density ridges defined by the tracers (see Appendix \ref{app:A} for a description). The final output of the algorithm is a collection of points. In Fig. \ref{fig:SCMS}, we show one example of the output of the SCMS algorithm in 2D space. The skeleton is extracted from a thin slice ($\sim 1.2$ Mpc thick) in region D6 at redshift $z=0$. The black dots represent the skeleton of the cosmic web. They are overimposed on the colour-coded KDE and the tracing substructures, which are marked with green dots. A thin cyan line separates the regions with $p(x) > \tau$, where $\tau=9.78\cdot10^{-6}$. The blue circle represents the virial radius of the massive clusters in the region. Note that some skeleton points lie inside their virial radius. A large part of the skeleton follows the densest regions and the tracer distribution. We notice, though, that there are some filaments that, while in the dense region, do not appear to be close to tracers. Most of them are very short (some are even single points), and since they are the result of the noise, at later stages, they will be removed.

\subsection{Sequential Chain Algorithm for Resolving Filaments}
    
    As shown and presented above, the outcome of the SCMS algorithm is a skeleton consisting of an ensemble of points that need to be automatically sorted and organized into filaments. For this purpose, we developed a Sequential Chain Algorithm for Resolving Filaments (SCARF).

    The SCARF algorithm is a two-step process: the initial step involves combining neighbouring points into a singular chain (see Sect. \ref{sec:SCA}), while the second step breaks down the single chain into separate filaments (see Sect. \ref{sec:RF}). A chain is a sequential arrangement of points where each point is next to its closest neighbour, forming a connected series like a linked chain. This two-step approach allows SCARF to organize and sort points of the skeleton into filaments in an effective and automated manner.

    \subsubsection{Constructing a single chain}
    \label{sec:SCA}
    
\begin{figure}
        \centering
        \includegraphics[width=0.95\linewidth]{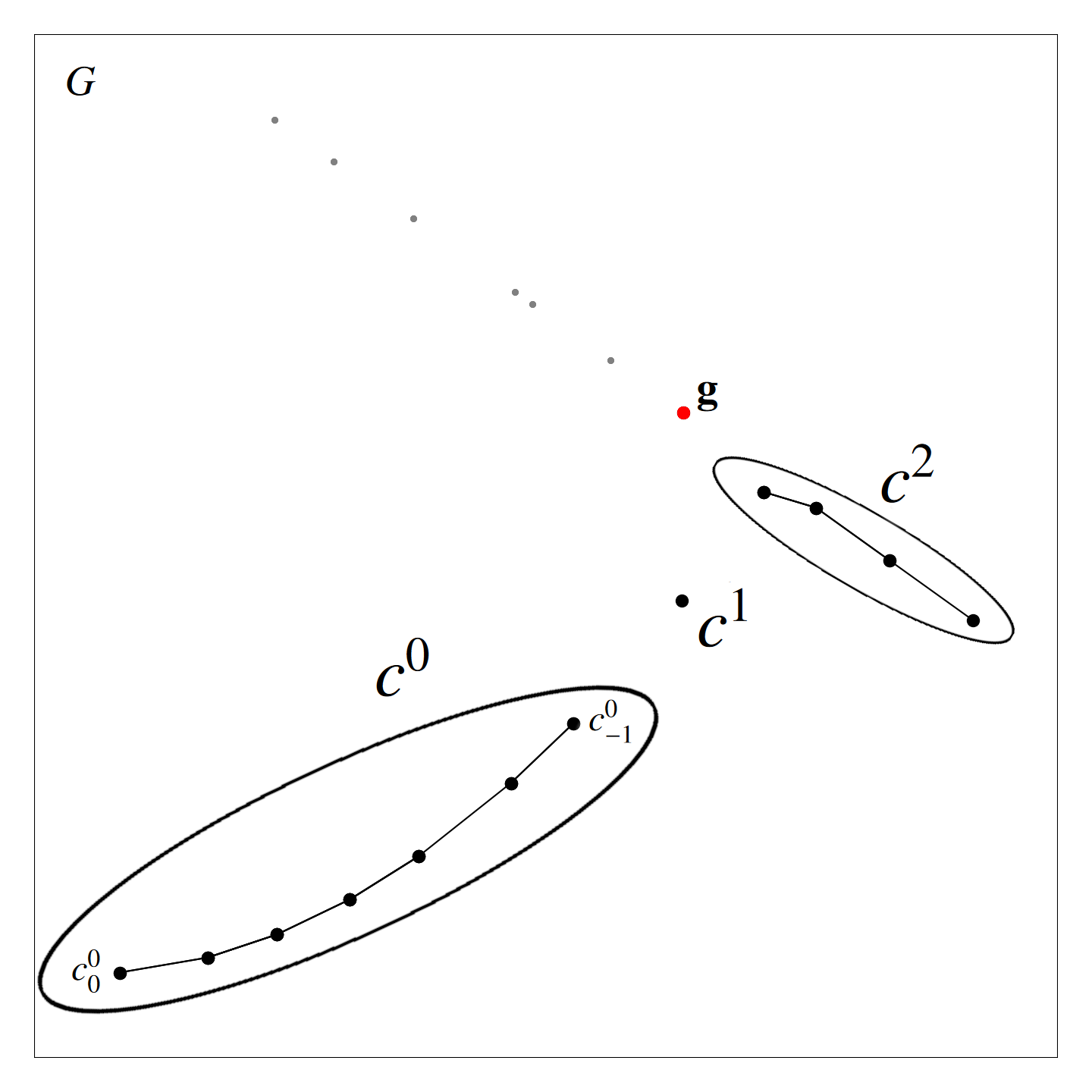}
        \caption{Visual presentation of an intermediate step of the first part of SCARF showing the key concepts on which the algorithm is based: the skeleton $G$ (all points), chains $c^{0}$, $c^{1}$ and $c^{2}$, chain ends $\mathbf{c}^{0}_{0}$ and $\mathbf{c}_{-1}^{0}$ and nearest neighbour $\mathbf{g}$ (red dot). The chain $c^{1}$ only has one point, which is fairly common when sorting points. Grey points are part of the skeleton $G$ but are yet to be sorted.}
        \label{fig:ScarfLabel}
    \end{figure}

    To facilitate the explanation, we start by presenting the main concepts used in the procedure, which are also displayed in Fig. \ref{fig:ScarfLabel}. Considering the entire collection of points forming the skeleton of the cosmic web or, in other words, the outcome of SCMS, the first step will be to identify the closest neighbour to the point of reference. This neighbour point is always referred to as $g$. To identify the chain, we will use the letter $c$. The final result of SCARF will be one unique chain for all points in the skeleton. However, at the intermediate steps (one of them is shown in Fig. \ref{fig:ScarfLabel}), more chains can be built, and these will be distinguished with a superscript indicating their numberings, e.g. $c^0$ and $c^1$ will be the first and the second chain that is created and so on. In each step, the initial and final points of the $i$-th chain are respectively called $c^i_0$ and $c^i_{-1}$. These points will change through time as the chain becomes longer. In some instances, the chain could be constituted of only one point and then $c^i_0\equiv c^i_{-1}$.

    The procedure starts by selecting a random point, which by definition is associated with the chain $c^0$. Subsequently, the closest companion, the point $g$, is identified, and its distance with respect to $c^0$ is computed. The closest companion to $g$ is searched, and if it is farther away than the distance between $g$ and $c^0$, then $g$ will be associated with the first chain $c^0$, which now constitutes of two points, and the search for another neighbour will continue. Otherwise, the point $g$ will not be connected to $c^0$, but it will be considered the first point of a second chain, $c^1$.

    Once two separate chains are identified, the procedure focuses on the closest neighbour of $c^1$, called again $g$. The program then compares the distance between the new $g$ and $c^1$ with the distance between $c^1$ and the end point of the chain $c^0$. If the latter is smaller than the former, then $c^1$ will be connected to $c^0$, and the two separate chains become one, named $c^0$. In the other case, $g$ will be attached to $c^1$ only if it does not have any other closest neighbours. In this circumstance, indeed, it will be the starting point of a third chain $c^2$.

    Generalizing the concept and the notation, whenever a new neighbour $g$ is found, the program computes its distance with its closest chain $c^i$ as the minimum distance between $g$ and the chain final points:
    \begin{equation}
        d_{g, c} = \min\{||\mathbf{g} - \mathbf{c}^{i}_{0}||, ||\mathbf{g} - \mathbf{c}^{i}_{-1}||\}.
    \end{equation}
    
    This distance will be compared with the distance between two previously identified subsequent chains, defined as the minimum distance between their final points:
    \begin{equation}
        d_{c, c} = \min\{||\mathbf{c}^{i}_{0} - \mathbf{c}^{i-1}_{0}||, ||\mathbf{c}^{i}_{0} - \mathbf{c}^{i-1}_{-1}||, ||\mathbf{c}^{i}_{-1} - \mathbf{c}^{i-1}_{0}||, ||\mathbf{c}^{i}_{-1} - \mathbf{c}^{i-1}_{-1}||\}.
    \end{equation}
    
    If $d_{c,c}$ is smaller, then the two chains are fused into one. Otherwise, the program checks whether $g$ has any other companion at a distance smaller than $d_{g,c}$. If it does not, then $g$ is associated with $c^i$; otherwise, $g$ will be the starting point of the new chain $c^{i+1}$.
    
    The procedure continues until all points from $G$ are connected in one single chain.

    \subsubsection{Resolving the filaments}
    \label{sec:RF}

    Once the entire skeleton is regrouped into a single chain, the second step of SCARF fragments the chain into individual filaments. The procedure starts by computing the vectors between two subsequent points: $\boldsymbol{\xi}_{i} = c_{i+1} - c_{i}$. Each vector has two main properties: length  $||\boldsymbol{\xi}||$ and direction, from which we can evaluate the change of direction of two neighbouring vectors as the angle
    \begin{equation}
        \theta_{i} = \arccos{(\boldsymbol{\xi}_{i} \cdot \boldsymbol{\xi}_{i-1})}.
        \label{eq:theta}
    \end{equation}
    
    The criteria used to break the chain into filaments are that either $||\boldsymbol{\xi}||>1$ Mpc or $\theta_i<\pi/6$. We select the first threshold to be of the order of the cluster radii, while the second limit is the minimum angle that still allows for smooth transitions of directions whenever the angle is below 30 degrees.

\begin{figure}
        \centering
        \includegraphics[width=0.95\linewidth]{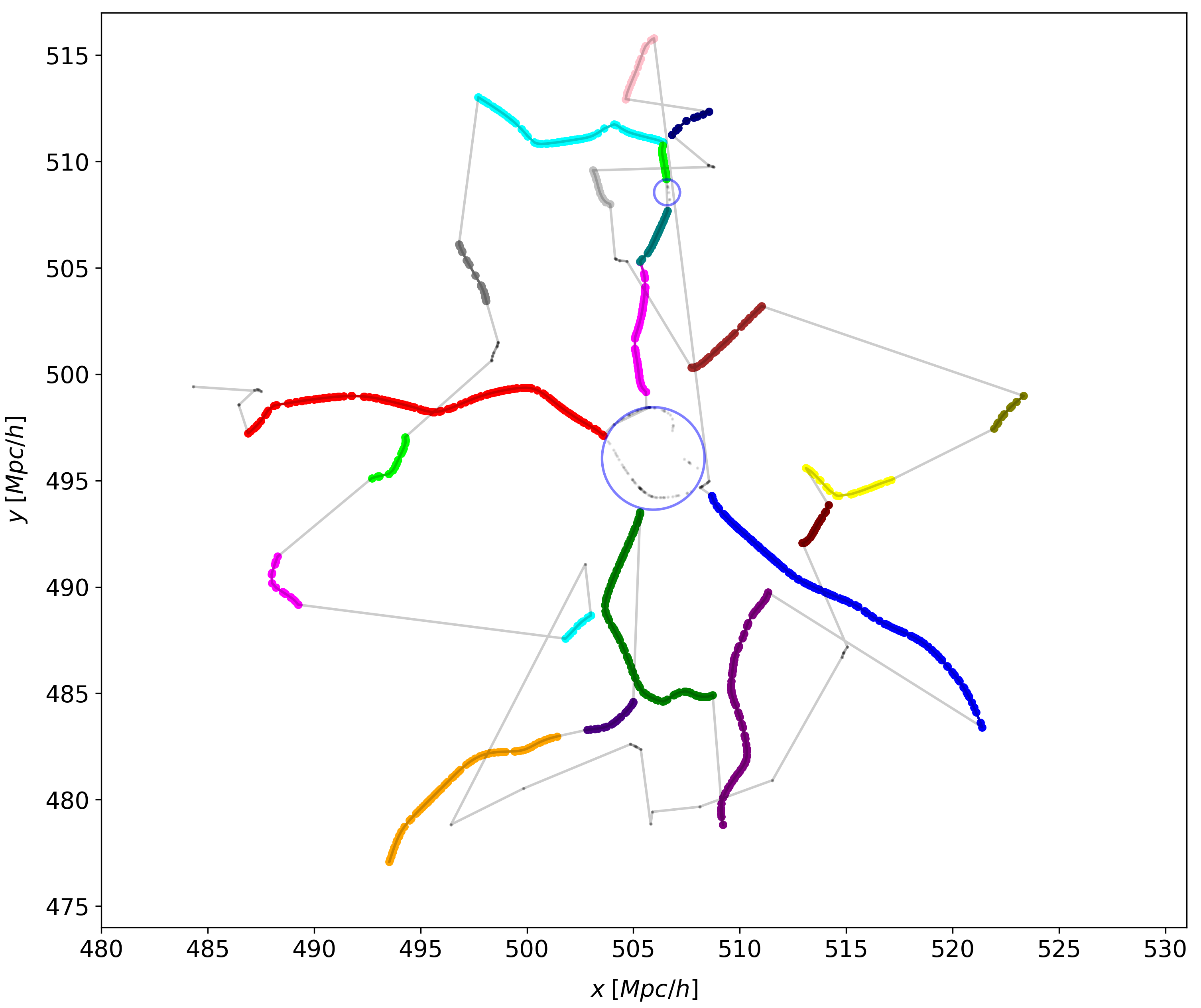}
        \caption{Final result of the SCARF algorithm applied to the same 2D region of Fig. \ref{fig:SCMS}. Different colours mark different filaments, while in a thin grey line, we show the chain identified in the first step of SCARF.}
        \label{fig:SCARF}
    \end{figure}

    In Fig. \ref{fig:SCARF}, we show the final results of the SCARF algorithm in 2D space where each filament is shown with a different colour. The points that lie inside the virial radius of the clusters with mass $M>10^{13} M_\odot$ (in this case, two) are removed. A low-opacity line represents a chain that connects all the skeleton points detected by the SCSM algorithm. Filaments arising from the noise are marked with small, low-opacity points. Because of their shortness, they are easily removed. In this particular 2D case, SCARF detected 21 filaments. The choices limiting the vector length and direction change angle are applied to all 3D skeletons for all the regions analyzed.
 
    \subsection{Mapping the filament}
    \label{sec:Map}
    
    A filament consists of $N$ points in successive order $f_0, f_1, ..., f_N$. We now presume that the filament is shaped like a bent cylinder. We denote a particle's position as $\mathbf{p}$. The longitudinal distance $l$ for a particle in $\mathbf{p}$ (e.g. gas or dark matter particle) is defined as the distance between the filament's starting point $f_0$ and the particle's projected point on the filament's spine and is calculated as 
    \begin{equation}
        l = \sum_{m=0}^n ||f_{m+1} - f_{m}|| +  \frac{(\mathbf{p} - f_{n}) \cdot (f_{n+1} - f_{n})}{||f_{n+1} - f_{n}||},   
    \end{equation}
    where $n$ is the sequence number of the point in the filament. Points $f_n$ and $f_{n+1}$ are the closest filaments' points to the projected point $\mathbf{p}$. The radial distance  $r$ is the shortest distance between the particle and the spine of the filament and is calculated as
    \begin{equation}
        r = \frac{||(\mathbf{p} - f_{n}) \times (f_{n+1} - f_{n})||}{||f_{n+1} - f_{n}||}.
    \end{equation}
    
\subsection{Length and radius of the filament}
\label{sec:Radius}

    We define the length of the filament as the sum of distances between neighbouring points:
    \begin{equation}
        L = \sum_{n=0}^N ||f_{n+1} - f_{n}||,
    \end{equation}
    where $N$ is the total number of points in the filament's spine. This differs from the usual definition of length, where filaments are connecting the nodes. SCMS does not define nodes, and this is the reason why we also remove points that lie inside the virial radii of groups and clusters, as described in Sect. \ref{sec:RF}. Consequently, the filaments may be slightly shorter than expected compared to other studies. Nonetheless, this definition still offers a reliable estimate of the filaments' geometrical properties.

    To estimate the filaments' radius, we first map the gas, dark matter and star particles surrounding the filament, as discussed in Sect. \ref{sec:Map}, up to $5$ Mpc from the filament spine. We obtained the particle distribution around the filament in $l$ and $r$ coordinate system. Our goal is to determine the filament's radius based on its overdensity, defined as
    \begin{equation}
        \delta = \frac{\rho}{\rho_{\rm{crit}}} - 1,
    \end{equation}

    where $\rho_{\rm crit}$ is the critical density\footnote{Critical density is computed using the cosmological parameters of the simulations as $ \rho_{\rm{crit}} (z) = \frac{3H_0^2}{8\pi G}  \bigg(\Omega_m(1+z)^3+\Omega_\Lambda\bigg).$} at the given redshift. This allows us to normalize the overdensity over all different regions, making them comparable. 

    We create an overdensity map that is described using the longitudinal and radial coordinates, $l$ and $r$. The map is divided into cells with a height and width of approximately $0.1$ Mpc. We map the particles at the same radial distance $r$ from the spine in the cell at the correspondent longitudinal distance $l$. We applied the Gaussian filter with $\sigma = 0.2$ Mpc, equivalent to the size of 2 neighbouring cells, to smooth the map and damp sharp irregularities. We then categorize the cells as overdense if their overdensity is $\delta > 0$ and underdense otherwise.

    Ideally, the filament would include most of the overdense cells close to the spine relative to the entire map. We can take two approaches: either determine a constant radius using the cells of the entire map or calculate a variable radius $R(l)$ using cells associated only with that $l$. In practice, since $R(l)$ can change rapidly, we applied a Gaussian filter with $\sigma = 1$ Mpc, allowing for a more gradual change in the filament's radius.

\begin{figure}
        \includegraphics[width=0.95\linewidth]{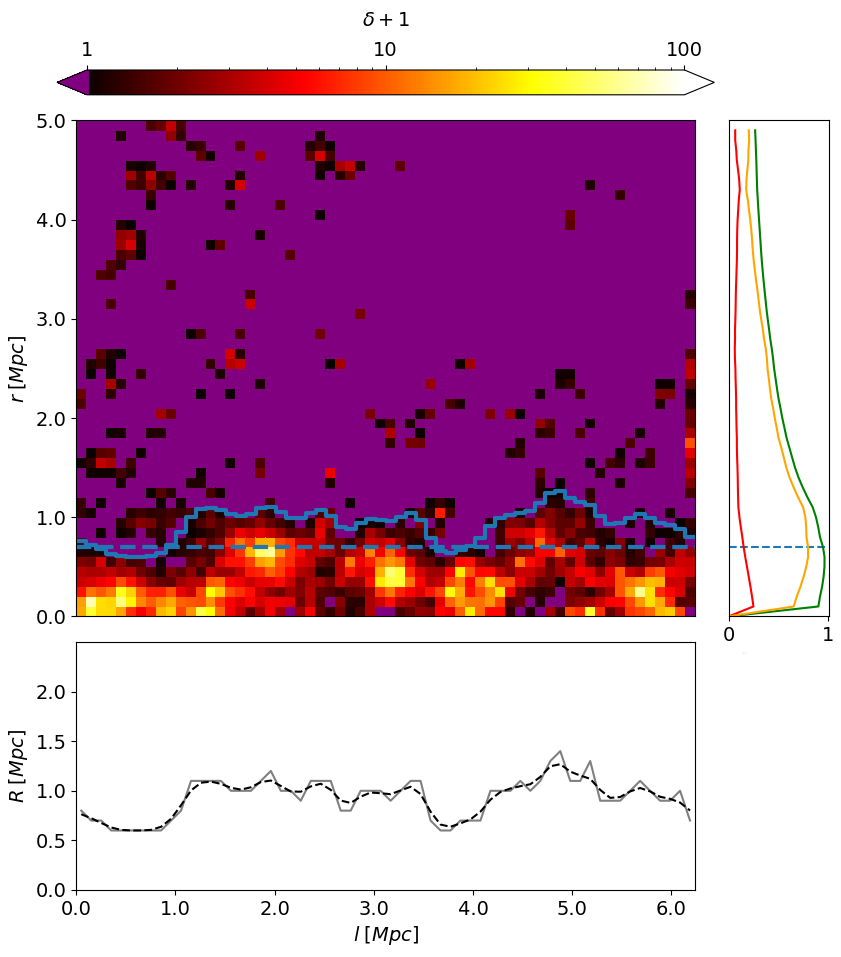}
        \caption{Map of overdensity, where the purple regions are below the selected overdensity threshold. The map also contains the computed constant radius (dashed blue line) and the variable radius $R(l)$ (solid blue line). On the right panel, at each $r$, we plotted the frequencies of overdense cells (red line for cells above that $r$ and green line for cells below $r$) and their difference (yellow line). The constant radius is shown as the radius where the difference in overdense cells is maximum. The variable radius $R(l)$ is plotted at the bottom of the figure. The grey solid line is the calculated radius, while the black dashed line is the radius with the applied Gaussian filter. The dark and red clumps at larger $r$ correspond to a larger overdensity of particles related to substructures. This map was obtained for one of the filaments in the D6 region (AGN simulation).}
        \label{fig:radius}
    \end{figure}

    The computation of both radii is shown in the following example. The main panel of Fig. \ref{fig:radius} illustrates an example of the colour-coded overdensity map around a single filament, with length $L\approx 6.2$ Mpc, in the D6 region (AGN simulation). The purple colour represents the underdense cells. The solid blue line shows the variable radius $R(l)$. We also evaluated a constant radius, indicated by the dashed blue line, in the following way. In the right panel, for each radius $r$ we computed the frequency of overdense cells for the upper area (red, above $r$) and lower area (green, below $r$). The constant radius is shown as the radius where the difference in the frequency of overdense cells is maximum (yellow line). In the lower panel, the calculated variable radius is represented by a solid line, while the dashed line illustrates the radius after applying the Gaussian filter. This example shows that the constant radius cuts off part of the overdense region, while the filament's variable radius traces better the filament's radial boundaries.

    However, issues can arise when no overdense cells are present at $l$, thus getting $R(l)=0$, or when objects, such as galaxies or small groups, are located outside the filament (red clumps in the overdensity map in Fig. \ref{fig:radius}). In that case, we can encounter a lone overdense cell (or a few) located far away from the filament's spine; because of that, the radius will extend beyond the actual filament. This can be mitigated by imposing a hard limit, where if the first few cells from the filament's spine are underdense, then the radius at that point is $R(l) = 0$. This overcomes, to some extent, the issue. In the cases where the average overdensity of the filament is below the threshold $\delta < 0$, we choose to remove the filaments from further analysis. On average, we had to remove 2 additional filaments per snapshot.
    
    \subsection{Physical properties of the filament}
    \label{subsec:phys_properties}

    To see if the filaments are the fair containers of cosmic baryons, we computed the gas, stellar and baryon depletion factors. They are calculated as in \cite{Planelles13}:
    \begin{equation}
        Y_{\rm{x}} = \frac{M_{\rm{x}}/M_{\rm{tot}}}{\Omega_{\rm{b}}/\Omega_{\rm{m}}},
        \label{eq:DepletionFactor}
    \end{equation}

    where $x$ can be gas, stars or baryons (gas and stars combined), $M_{\rm{x}}$ is the total mass of the x component, $M_{\rm{tot}}$ total mass of the filament, and $\Omega_{\rm{b}}$ and $\Omega_{\rm{m}}$ the density parameters used in the simulations.
        
\begin{figure*}
        \includegraphics[width=0.95\linewidth]{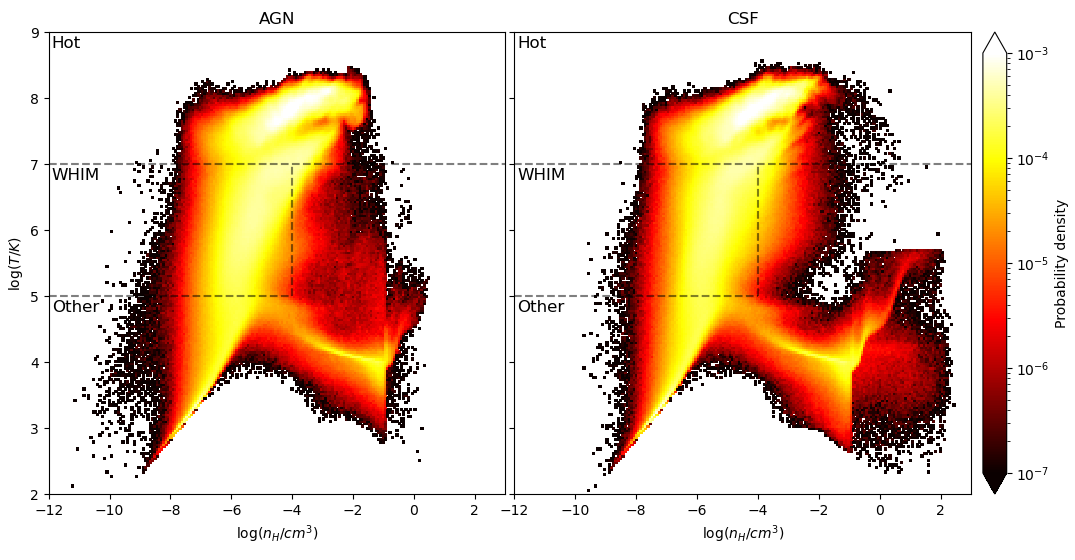}
        \caption{Phase diagram of hydrogen number density $n_H$ ($x$-axis) and temperature $T$ ($y$-axis) for all gas particles in all regions at redshift $z=0$ in AGN (left) and CSF (right) simulations. The dashed lines mark the limits for each gas phase defined in the text (Hot/WHIM/other). The colour code (right) refers to the probability density. 
        }
        \label{fig:PhaseDiagram}
    \end{figure*}

    We analysed the properties of the gas included in a larger region around the spine of the detected filaments. For the purpose of this work, we studied the gas components, dividing them into three phases based on temperature and hydrogen number density $n_H$, as done, for example, in \cite{2019MNRAS.486.3766M}. The hot gas phase corresponds to gas having temperatures above T $> 10^7$ K and any number density. This phase is common in the potential wells of massive clusters, but as we will see, we can find this phase in filaments as well, especially near the filament ends (if they are near a cluster). For the WHIM gas phase, we used the range $10^5 \rm{K} < $ T $< 10^7$ K for temperature and $n_{\rm{H}} > 10^{-4}(1+\rm{z})$ cm$^{-3}$ for the hydrogen number density. We expect most of the gas to be found in this phase. The remaining gas is treated separately and includes the colder gas associated with the ISM of galaxies and cold IGM, as well as the warm CGM created by shock heating and feedback processes near galaxies. In Figure \ref{fig:PhaseDiagram} we plotted the phase diagram of hydrogen number density $n_H$ and temperature $T$ for all gas particles in all regions combined at redshift $z=0$ in AGN and CSF simulations. Dashed lines separate the different gas phases that we have defined previously: hot, WHIM and other. Here, we can also note the difference between AGN and CSF simulations. The biggest difference is seen at the tail-end of the diagram, where high-density and low-temperature gas is located. This is the star-forming gas, which is more abundant in CSF simulations, while it is efficiently removed from the star-forming phase by AGN feedback.

    To show the different behaviour of each gas component inside the filamentary structures, we calculated the mass fraction for each gas phase as $M_{\rm x}/M_{\rm{gas}}$, where $\rm x$ is the mass fraction of the desired gas phase and M$_{\rm{gas}}$ the total gas mass inside the filament.

    To explore the metallicity content of filaments, we obtained the iron metallicity $Z_{Fe}$ for each gas phase in the filament as
    \begin{equation}
        Z_{\rm{Fe}} = \frac{\sum_i m_{\rm{Fe}, i}}{\sum_i m_{\rm{H}, i}},
    \end{equation}
    where the sum is done over all the particles contained in the region of the filament and where the solar abundance is $Z_{Fe, \odot} = 1.77 \cdot 10^{-3}$ as in \cite{GS98}. To be consistent with the observed temperatures, we also used the definition of spectroscopic-like temperature $\rm{T}_{\rm{SL}}$, as described in \citet{Mazzota04} \citep[see also][]{Rasia05}
    \begin{equation}
        T_{\rm{SL}} = \frac{\sum_i \rho_{\rm{i}}  T_{\rm{i}}^{0.25} m_{\rm{i}}}{\sum_i \rho_{\rm{i}} T_{\rm{i}}^{-0.75}  m_{\rm{i}}},
        \label{eq:T_SL}
    \end{equation}
    where $T_{\rm{i}}$ is the temperature of the gas particle, $\rho_{\rm{i}}$ its density and $m_{\rm{i}}$ its mass.
    
    \section{Results}
    \label{sec:Results}

    First, we applied the SCMS algorithm on the selected set of the Dianoga regions, obtained with two different physics (AGN and CSF), as described in Sect. \ref{sub:dianoga}. To study the evolution of gas properties, we selected redshifts $z = 0, 0.25, 0.51, 0.76, 1, 1.26$ and $1.48$. We removed all the points of the skeleton that fall inside the virial radii of the groups and clusters with masses above $10^{13} M_{\odot}$. After that, we ran the SCARF algorithm on the remaining skeleton. The final output is one catalogue of filaments per region, redshift and physical model.

    We calculated the physical length $L$ of all the filaments as specified in Sect. \ref{sec:Radius}. We only keep filaments with length $L>2$ Mpc for further analysis. Shorter filaments could be, at best, considered galaxy bridges and are entirely disregarded. With this, we also remove the false filaments, as discussed in Sect. \ref{sec:SCMS}. Almost all of the detected filaments are inside the high-resolution regions entirely, as they are either connecting two massive halos within this denser region or are situated in between filaments. A few of the filaments do not have a cluster or group at one of the ends. We still consider these filaments, as our main focus is the study of diffuse gas.
    
\subsection{Geometrical properties}

    Before analysing the geometrical properties of filaments, we provide in Table \ref{tab:FilamentStats} the number of filaments with $L>2$ Mpc found per region, redshift and simulation type. We do expect that the number of filaments is related to the volume of the region. However, we see a steady increase in time of the filament absolute number in all regions, the numbers growing by $\sim 60\%$ for the smallest ("isolated") regions, D5 and D9, and $\sim 40\%$ for the other three largest regions. This is expected since the density contrast increases with time, especially in isolated regions \citep[see, e.g.][]{Galarraga24}. We checked that, in general, filaments become longer with lower redshift in proper coordinates, though not considerably. There is no significant difference in the evolution of the number of filaments between AGN and CSF simulations. We do observe slight deviations between AGN and CSF simulations, but the difference is, in most cases, below $5\%$. Since the formation of the cosmic web is mostly determined by gravity, a similar number of filaments is expected as physical models used in AGN and CSF simulations mostly affect the gas component.
    
    \begin{table}
    	\centering
    	\caption{Number of filaments detected per region, redshift and simulation type. All filaments have a length $L$ of at least $2$ Mpc.}
    	\label{tab:FilamentStats}
    	\begin{tabular}{l|ccccc}
    		\hline
    		z & D1 & D5 & D6 & D9 & D22 \\
            \hline
            \multicolumn{6}{c}{AGN} \\
    		\hline
                1.48 & 154 & 67 & 192 & 75 & 344 \\
                1.26 & 174 & 92 & 213 & 94 & 370 \\
                1.00 & 183 & 107 & 256 & 100 & 410 \\
                0.76 & 201 & 123 & 278 & 119 & 455 \\
                0.51 & 197 & 130 & 295 & 149 & 481 \\
                0.25 & 191 & 147 & 311 & 172 & 543 \\
                0.00 & 211 & 170 & 323 & 183 & 564 \\
            \hline
            \multicolumn{6}{c}{CSF} \\
    		\hline       
                1.48 & 152 & 82 & 191 & 72 & 348 \\
                1.26 & 162 & 83 & 227 & 92 & 373 \\
                1.00 & 190 & 108 & 251 & 109 & 411 \\
                0.76 & 195 & 127 & 275 & 114 & 459 \\
                0.51 & 186 & 130 & 284 & 135 & 500 \\
                0.25 & 192 & 143 & 292 & 169 & 515 \\
                0.00 & 214 & 156 & 326 & 189 & 547 \\
    		\hline
    	\end{tabular}
    \end{table}
    
    Along with the filaments' length, we can also observe their shape (whether the filament is straight or curved). Therefore, we computed the distance $D$ between the filament's first and last point and compared it with the filament length $L$.

\begin{figure}
        \includegraphics[width=0.95\linewidth]{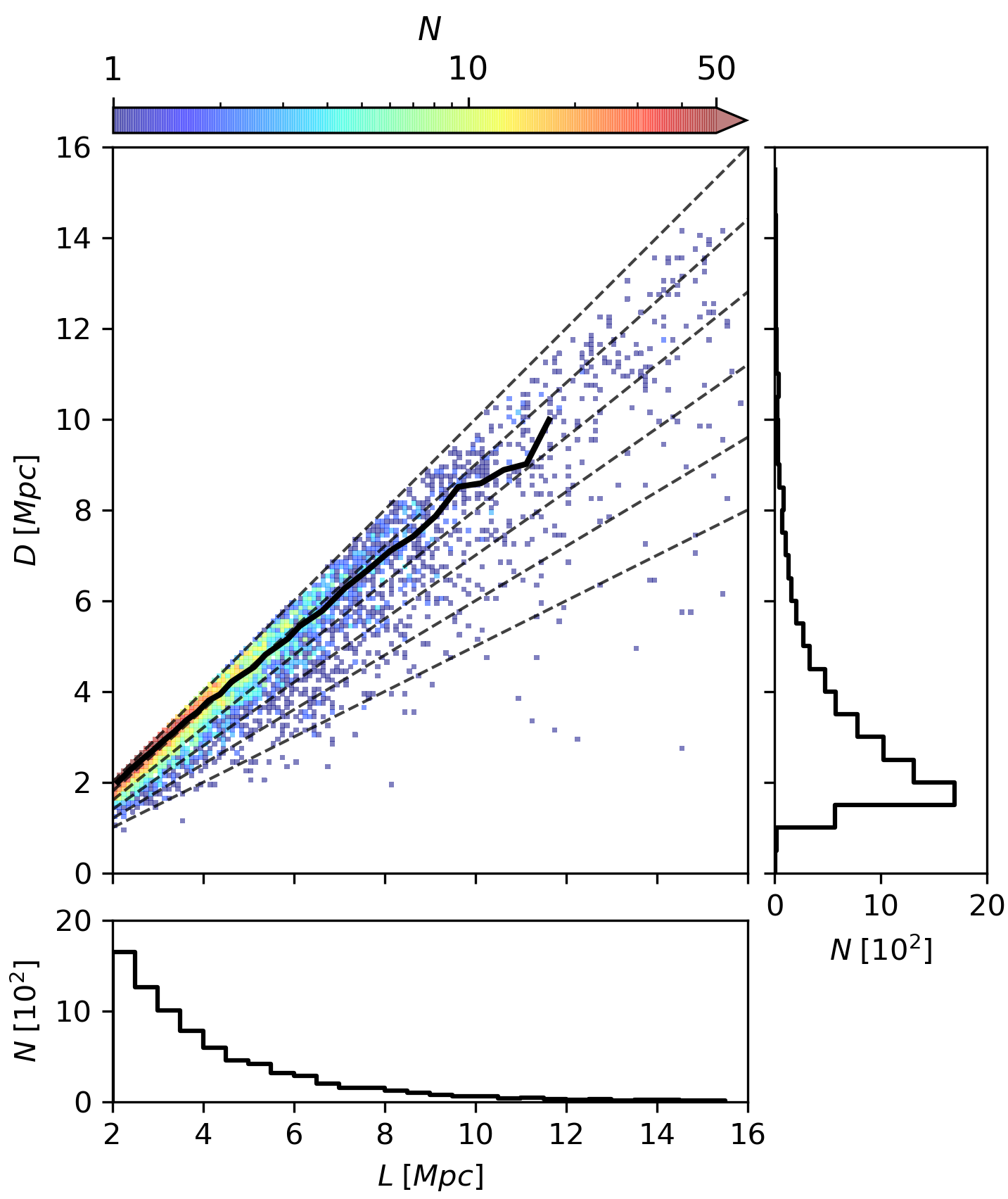}
        \caption{A 2D histogram between $L$ and $D$, with $N$ as the color-coded number of counts. Dashed lines represent the fraction $D/L$ with coefficients of $1$, $0.9$, $0.8$, $0.7$, $0.6$ and $0.5$ from top to bottom, respectively. The black line represents the median value of $D$ at a given length $L$. On the sides are histograms of $L$ and $D$. This is for all detected filaments across all regions, redshifts and simulations with $L\geq2$ Mpc. This is a distribution for AGN simulations, but the distribution for CSF is almost identical.}
        \label{fig:FilamentStats}
    \end{figure}

    Since, in this case, we are not interested in environmental, model or redshift dependency, we combined all filaments with $L\geq2$ Mpc from all regions and redshifts. In Fig. \ref{fig:FilamentStats}, the 2D histogram between $D$ and $L$ is plotted, with dashed lines representing the values of $D/L$ and solid black line representing the median values of $D$ at given $L$. On the side are the histograms for $L$ and $D$. The ratio $D/L$ represents the shape of the filament; the lower the value, the more curved the filament. As expected, since we are working with a low-volume simulation box, most of the filaments have lengths below $9$ Mpc, with number counts falling with length. Similarly, this happens with distance $D$ since this quantity is correlated with the length of the filament. According to multiple works \citep{Galarraga21, Galarraga24, Wang24}, filament populations can be divided into short and long filaments. Though there is no universally set limit, in all of the works, long filaments have a length of at least $20$ Mpc. This means that in this work, we exclusively study short filaments. The median line (solid black line in Fig.\ref{fig:FilamentStats}) indicates that the longer the filament, the more likely it is to be curved. The median value is close to $D/L=0.9$ for short filaments. With larger lengths, the median value is moving towards the lines with lower values for $D/L$, indicating that the longer filaments are more likely to be curved since they do interact tidally with clusters located close to the filaments \citep{Colberg05, Gonzales10, Cautun14}. Even so, there are a number of longer filaments that are straight and smaller filaments that are curved.
    
\begin{figure}
        \includegraphics[width=0.95\linewidth]{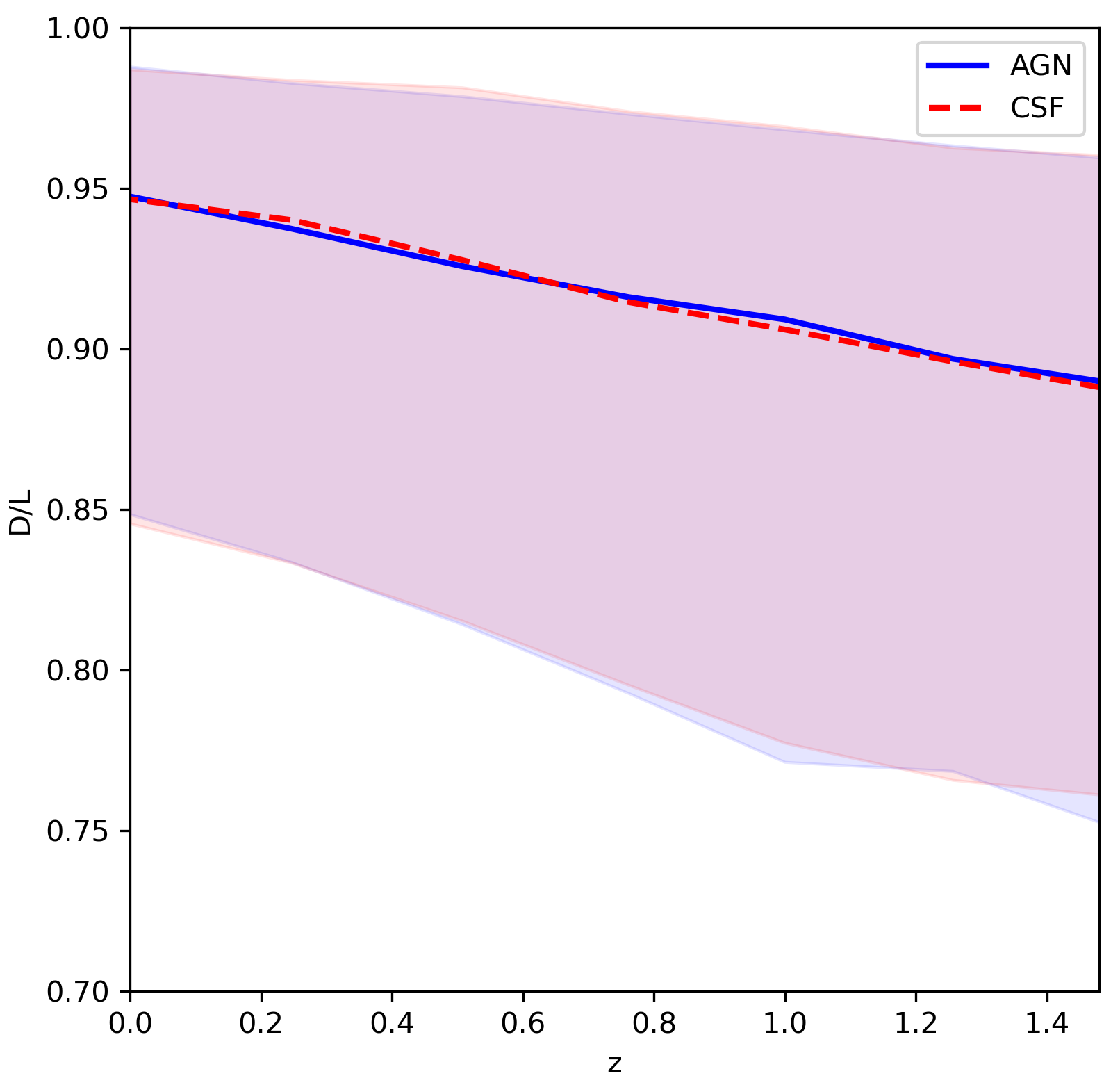}
        \caption{Median value of shape $D/L$ of the detected filaments in all regions in dependence to redshift $z$, for AGN (solid blue) and CSF (dashed red) simulations. The shaded regions are values between $16^{th}$ and $84^{th}$ percentile.}
        \label{fig:Shape}
    \end{figure}

    We explored how the shape of the filaments changes with redshifts. Shown in Fig. \ref{fig:Shape} are median values plotted against redshift $z$ for AFN and CSF simulations. The median values were calculated at each redshift combined for all the filaments in all regions. We can observe the filaments' shape is consistently rising with lower redshift, which means that the filaments become straighter with time. A similar conclusion was reached in \cite{Cautun14}, where the filament's shape was similarly defined. As expected, the evolution to straighter filaments is similar for both AGN and CSF simulations. 

\subsection{Global properties}
\label{sec:PhyProp}

\begin{figure}
        \includegraphics[width=0.95\linewidth]{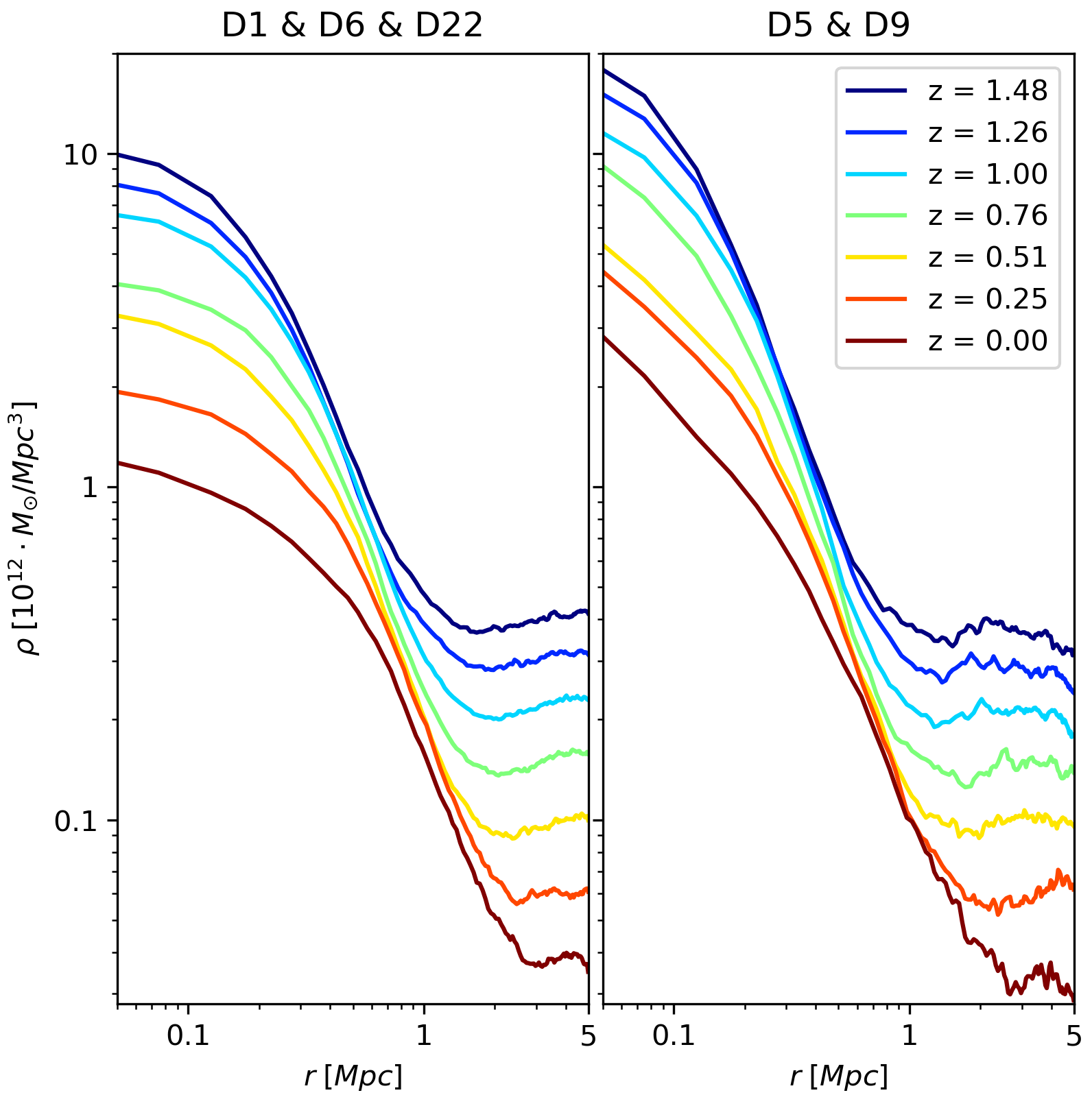}
        \caption{Mean radial profiles of dark matter for all filaments at different redshifts. The left graph has filaments from regions D1, D6 and D22, while the right has filaments from regions D5 and D9. These radial profiles were extracted from AGN simulations but are the same in CSF simulations. The radius and density are both in proper units.}
        \label{fig:Radial}
    \end{figure}

    We calculated the radial profile for each filament and then averaged all of them into a single radial profile at a given redshift, the same way as in \cite{Galarraga20, Galarraga24}. The results are shown in Fig. \ref{fig:Radial}, where we have plotted the dark matter radial profiles for all filaments from regions D1, D6 and D22 (left panel) and all filaments from regions D5 and D9 (right panel), separately. These radial profiles are in agreement with the results from other works \citep[e.g.][]{Tanimura20may, Galarraga20, Galarraga24, Wang24}. For all of the regions, as the redshift decreases, we observe a decline in physical density. It is also clear that the filaments around the more isolated objects have a higher density close to their spine. Still, they fall quicker to values similar to the other three regions at larger radii. In the same figure, we also observe the widening of radial profiles with decreasing redshift. This indicates that, on average, the radius of filaments increases in time. This aligns with the findings in \cite{Wang24}, where they examined the radii of short, medium, and long filaments in the MilleniumTNG simulations. For short filaments (median value $7.56$ Mpc), which most of the filaments in our study are, the radius decreased until redshift $z=2$ and then began to increase. Similar trends were observed for medium and long filaments, with the turning point occurring at redshift $z=1$. However, for long filaments, the radius remained relatively unchanged after this redshift. 
    
    The following computations are performed for all identified filaments across five regions at seven distinct redshifts. We begin by calculating for each filament the mean radius, denoted as $\langle R \rangle$, by taking the mean of the variable radii. We verified that, similarly to what is shown in Fig.\ref{fig:Radial}, $\langle R \rangle$ increases with redshift. Subsequently, we gather all particles located within the filaments' variable radius and collect their masses to obtain the total mass of the filament. Given that we know the filament's mass and volume, we can compute its average density $\rho$ and its average overdensity $\delta$. 
    
    \begin{figure*}
        \includegraphics[width=0.95\linewidth]{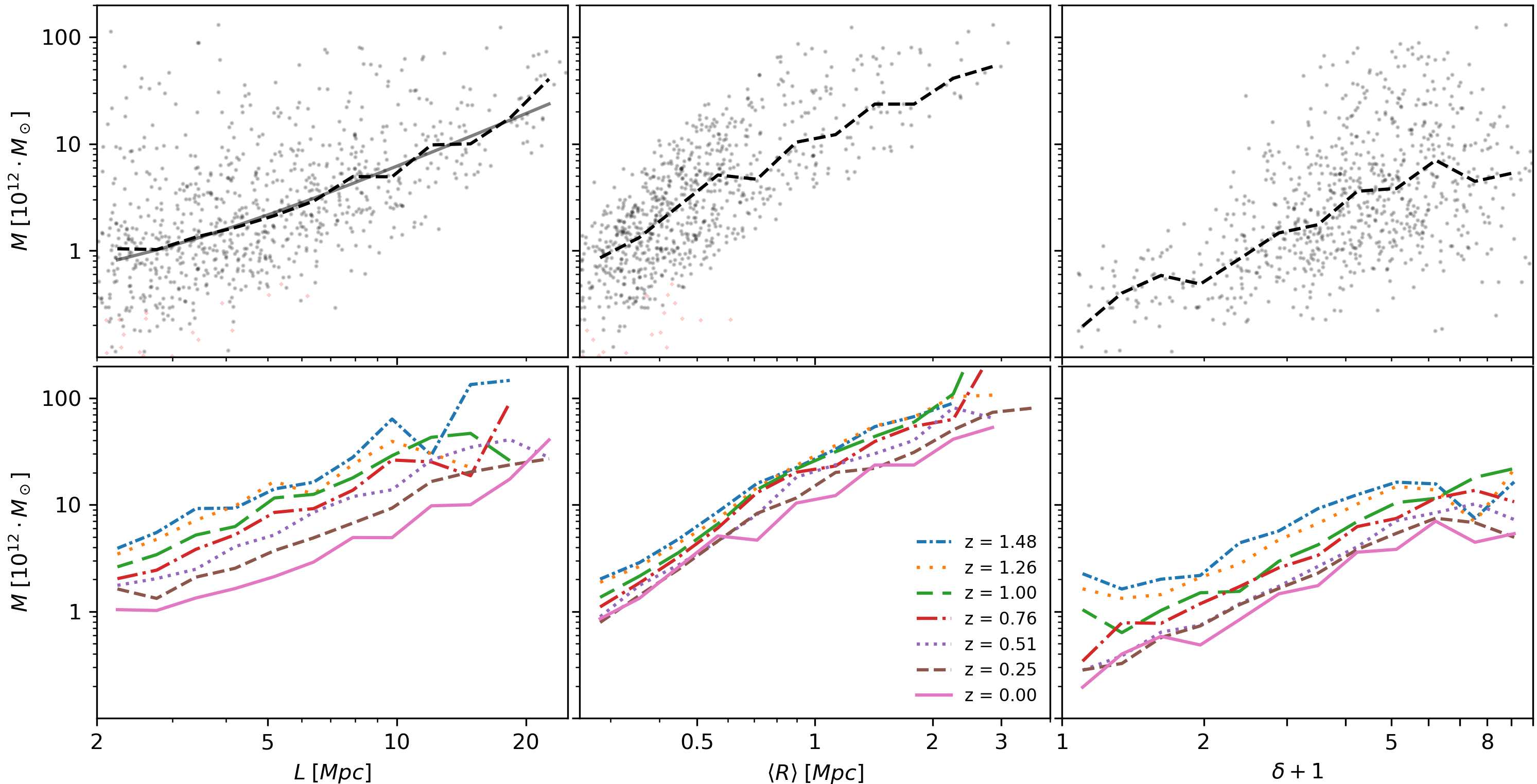}
        \caption{Top panels: The distribution of mass of the filament M versus its length L (left), mean radius $\langle R \rangle$ (middle) and overdensity $\delta + 1$ (right) at redshift $z=0$. A dashed curve represents a median mass at a given filament range in x-axes. A fitted power law curve is plotted on a mass versus length graph, with a solid line representing the relation M $\propto$ L$^{1.7}$. The red dots represent the discarded filaments with $\delta < 0$. Bottom panels: The median mass values of filaments with respect to their length L (left), mean radius $\langle R \rangle$ (middle) and overdensity $\delta + 1$ (right) at different redshifts. For redshifts, the different colours and lines are reported in the legend (middle plot). All the filaments on the plot or from the AGN simulation.}
        \label{fig:MassVsLength}
    \end{figure*}
    
    In Fig. \ref{fig:MassVsLength}, the top-left panel displays the distribution of total filament mass against its length for filaments at redshift $z=0$ in AGN simulations. The results for CSF simulations are similar. The graph includes median mass values within specific length ranges, revealing a consistent increase in mass. A power-law fit indicates a relation of M $\propto$ L$^{1.7}$. \cite{Cautun14} determined a similar power-law relation, $\propto$ L$^{2.2}$, based on filaments in Millennium simulations detected with a different algorithm, NEXUS$+$. Notably, their simulations featured longer and, consequently, more massive filaments, a larger sample size, and exclusively dark matter-filled filaments. The existence of this correlation spanning a different range of lengths and masses implies that the longest filaments are not a fortunate connection of short filaments but rather clearly defined and possibly persistent structures. The bottom-left panel illustrates median lines for seven redshifts. Filaments consistently adhere to the same mass-length relation but exhibit decreasing mass with lower redshifts. 

    Our results might seem to go against the expectation that filaments become thinner as time passes, becoming more massive and diminishing radii with redshift. However, we should keep in mind that we are not studying large-scale filaments from cosmological boxes but relatively shorter filaments in denser regions populated with clusters. This aspect is likely influencing the evolution of filaments. Massive clusters will grow thanks to the mass flow from filaments, but we expect the filaments will not accrete mass from surroundings at the same rate. This would require a deeper investigation that is out of the scope of this paper. Moreover, as noted also in \cite{Cautun14}, the correlation between the width and density of filaments depends on evolutionary processes. They found that thin filaments are usually found in underdense regions (areas where large-scale filaments form and persist), while thick ones are in overdense areas. The area we are studying is rather dense and populated with a large number of clusters that act as gravitational attractors.
    
    A similar plot is presented for mass versus mean radius in the middle panels of Fig. \ref{fig:MassVsLength}. Most of the mean radius are below $1$ Mpc, which is a standard radius for the filaments, most recently determined by \cite{Wang24}. Still, the trend indicates that larger mean radii correspond to more massive filaments, which also correspond to longer filaments. Additionally, the radius of the filament appears to increase with redshift at the same mass, which is consistent with Fig. \ref{fig:Radial}. The panels on the right depict mass versus overdensity. Initially, the mass appears constant at lower overdensity but rises steadily with overdensity until it becomes nearly constant again. Note that, as seen in the upper right panel, most of the filaments have overdensity larger than $\delta > 2$.
 
\subsection{WHIM and hot gas phase properties}
    To understand the broader evolution of the mass in filaments, we computed the fractions of the main matter components: dark matter, gas and stars within the boundary of the variable radius $R(l)$. We noticed that the filaments located at the borders of the high-resolution regions have a very low number of gas particles. Therefore, we removed those with $\frac{M_{\rm{gas}}}{M_{\rm{tot}}} < 0.05$ from further analysis. We then calculated the depletion factor for gas, stars and baryons using Eq. \ref{eq:DepletionFactor} for both AGN and CSF simulations. The results at redshift $z=0$ are shown in Table \ref{tab:Depletion}, where we can notice that filaments are not fair containers of the baryons in both simulations since $Y_{baryons}$ is around $\sim 0.79$ in AGN and $\sim 0.83$ in CSF simulations, lower than unity. This is broadly consistent with the result by \cite{Galarraga22}, where they observed short filaments in IllustrisTNG simulations having $Y_{baryons} \sim 0.9$. They concluded that there is a baryon deficiency at $r<0.7$ Mpc, which coincides with the typical radius of the filaments we study. The gas in the filaments within AGN simulations has a comparable fraction with that of the CSF simulations. In turn, $Y_{\rm stars}$ in CSF filaments is almost twice the value of that in AGN simulations, having both a large error. This is expected since CSF simulations are characterized by overcooling and missing the AGN feedback needed to regulate the star formation (e.g. AGN feedback), and overdense regions have higher production of stars. This increases the abundance of stars in the filaments in CSF simulations while simultaneously reducing their gas abundance. On the other hand, the reduced star formation in the AGN simulations makes more baryons available to be part of the diffuse medium, but such extra baryons are, in fact, displaced by the same AGN feedback. As a result, while $Y_{stars}$ in AGN simulations decreases, there is almost a compensation of the two effects in determining the value of $Y_{gas}$. We can notice that at the end the overall baryon content of filaments is slightly reduced in AGN simulations.

    \begin{table}
        \centering
        \caption{Depletion factors of gas, stars and baryons for filaments. in AGN and CSF simulations at redshift $z=0$, with error values representing the standard deviation.}
        \begin{tabular}{l|cc}
            & AGN & CSF\\
            \hline
           $Y_{\rm gas}$     & $0.733 \pm 0.202$ & $0.727 \pm 0.188$ \\
           $Y_{\rm stars}$   & $0.058 \pm 0.046$ & $0.105 \pm 0.085$ \\
           $Y_{\rm baryons}$ & $0.792 \pm 0.248$ & $0.832 \pm 0.274$ \\
        \end{tabular}
        \label{tab:Depletion}
    \end{table}
    
\begin{figure}
        \includegraphics[width=0.95\linewidth]{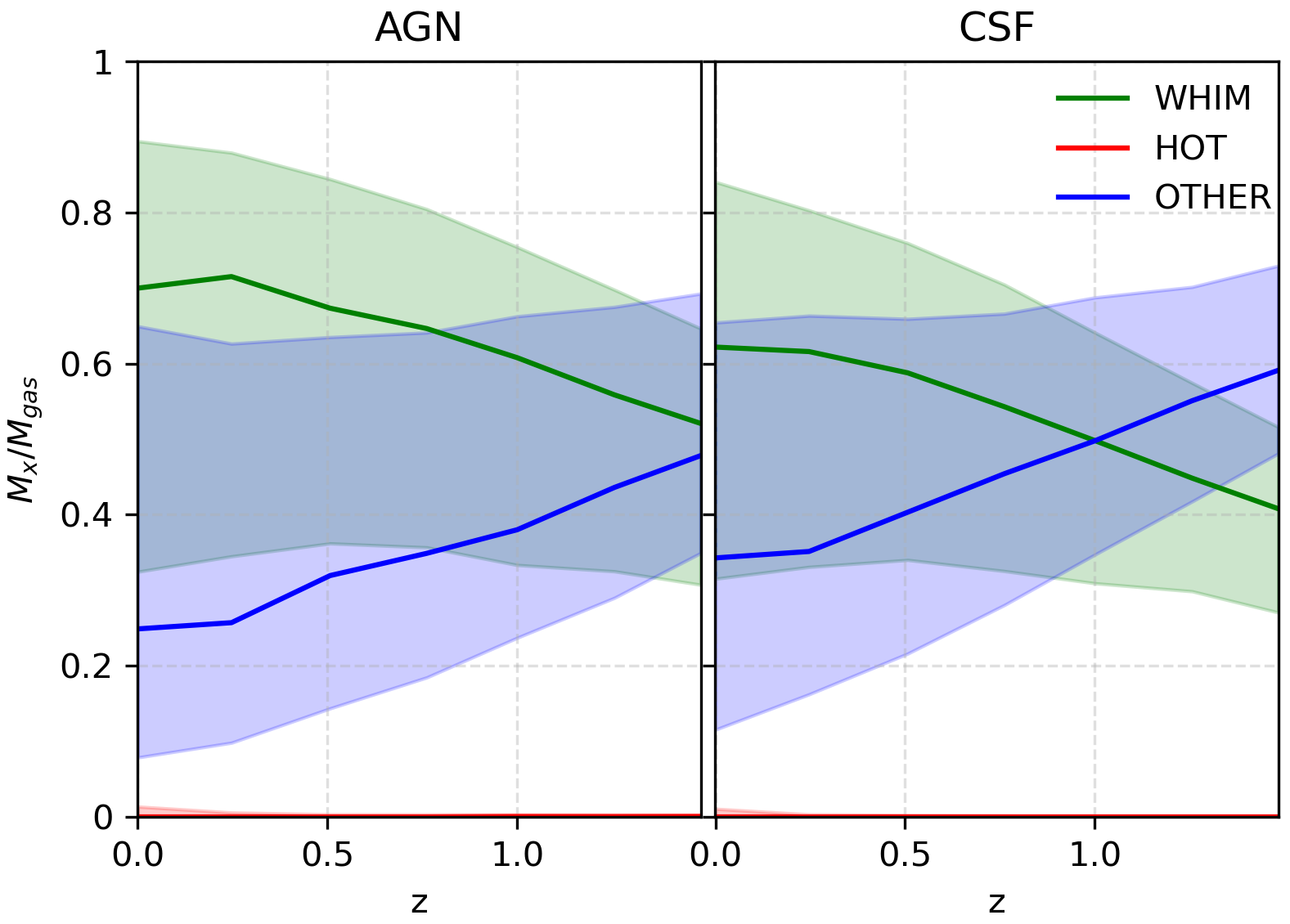}
        \caption{Median values (solid lines) of gas mass fractions where x is WHIM, hot and other gas phases. The left panel shows results for AGN simulations and the right panel results for CSF simulations.}
        \label{fig:GasStatistics}
    \end{figure}

    We checked that the mean mass fractions of baryons and dark matter do not change much in time (around $5\% $ in AGN and $\% 10$ in CSF simulations with respect to the present day value). The most notable rise of the mass fraction is for the stars in the CSF simulations, which is expected due to the high star-formation rate. Because of the relatively small mass fraction in stars, this does not visibly impact the gas and dark matter mass fraction. We then calculated the mass fractions of the three gas phases, defined in Sect. \ref{subsec:phys_properties} and calculated the median mass fraction for each gas phase and the $16^{\rm{th}}$ and $84^{\rm{th}}$ percentiles for AGN and CSF simulations at each redshift. The results are shown in Fig. \ref{fig:GasStatistics}. We immediately notice that the gas phases follow slightly different evolutions depending on the simulation. This is broadly consistent with what was found by \cite{Tornatore2010}, where slightly different simulations were analysed in a broader context of the WHIM content inside smaller cosmological boxes. In AGN simulation, the WHIM gas phase consistently stands out as the most abundant across different redshifts, while the hot gas phase remains at very low levels. We recall that the filaments in this work are outside the virial radii of groups and clusters. Notably, the WHIM gas phase becomes even more abundant as redshift decreases. At redshift $z = 0$, its median mass fraction reaches almost $0.6-0.7$ in the two physical models. In similar works, this value is between $0.5-0.7$ \citep{2019MNRAS.486.3766M, Galarraga21}. Meanwhile, the presence of other gas phases gradually decreases. Though the abundance of the hot gas phase is barely noticeable, it does increase slightly at lower redshift. This indicates that the gas temperature in the filaments is slowly rising with lower redshift, likely because of the shock heating \citep{2019MNRAS.486.3766M}. Due to that, the filaments would possibly be easier to detect with X-ray telescopes at lower redshifts. The two models show very similar trends with redshift. However, the fractions of the WHIM and the colder gas reach equality earlier for the AGN simulation.

    \begin{figure}
        \includegraphics[width=0.95\linewidth]{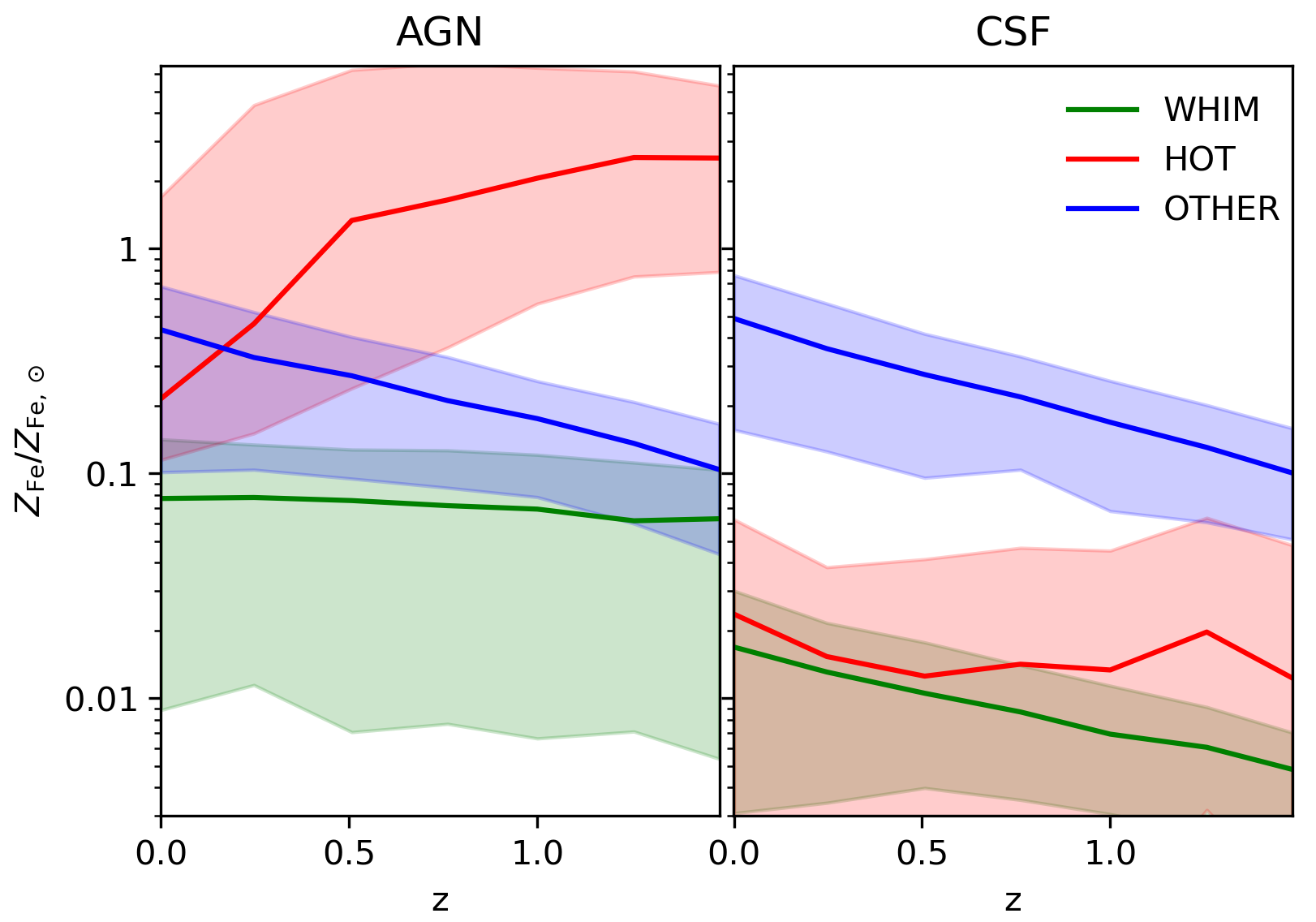}
        \caption{Median metallicity $Z_{Fe}$ for WHIM, hot and other gas phases in redshift dependence. The left panel shows results for AGN simulations and the right panel results for CSF simulations.}
        \label{fig:Metallicity}
    \end{figure}

    One of the most striking differences between the two models can be seen in the metal content of the filaments. Furthermore, metallicity will be important in the observational detection of filaments. Therefore, we calculate the metallicity $Z_{Fe}$ for each gas phase in every filament, as well as the median values at each redshift. The results are shown in Fig. \ref{fig:Metallicity} for AGN and CSF simulations, respectively. 
    
    In AGN simulations, the hot gas phase exhibits the highest metallicity among the three considered phases, with the other gas phases ranking second and the WHIM displaying the lowest metallicity. Over time, the metallicity of the hot gas gradually decreases, while both the WHIM and other gas phases experience a slow but steady rise in metallicity. At redshift $z=0$, the metallicity of the hot gas phase falls below the value of other gas phases. These results align with those of \cite{2019MNRAS.486.3766M}, who found that the decrease in metallicity at lower redshifts is attributed to increasing temperatures caused by shock heating. This results in the transition of the metal-poor WHIM phase (as well as the warm CGM phase) to the hot phase. Since hot gas is not abundant in the filaments, as shown in Fig. \ref{fig:GasStatistics}, the addition of metal-poor gas decreases the median metallicity.

    Conversely, the hot gas phase in CSF simulations shows much lower metallicity than in AGN simulations. It has barely larger values than WHIM, which also has lower metallicity than in AGN simulations. Other gas phases, though, exhibit similar values at all redshifts in the two physical models. Notably, across all three gas phases in CSF simulations, there is a collective increase in metallicity over time, marking a distinctive difference from the trend observed in AGN simulations, likely due to a higher star-formation rate, resulting in larger metal production. The overall metallicity values are higher in AGN simulations compared to CSF simulations, highlighting the effect of the AGN feedback model in simulations. This is consistent with the previous results shown in \cite{Biffi18} where the AGN feedback, especially active at $z=2$, ejected metal-enriched gas from the small potential well of high-redshift galaxies. It is, therefore, expected that in AGN simulations, filaments retain gas with high metallicity, funnelling it toward clusters.

    \begin{figure*}
        \includegraphics[width=0.95\linewidth]{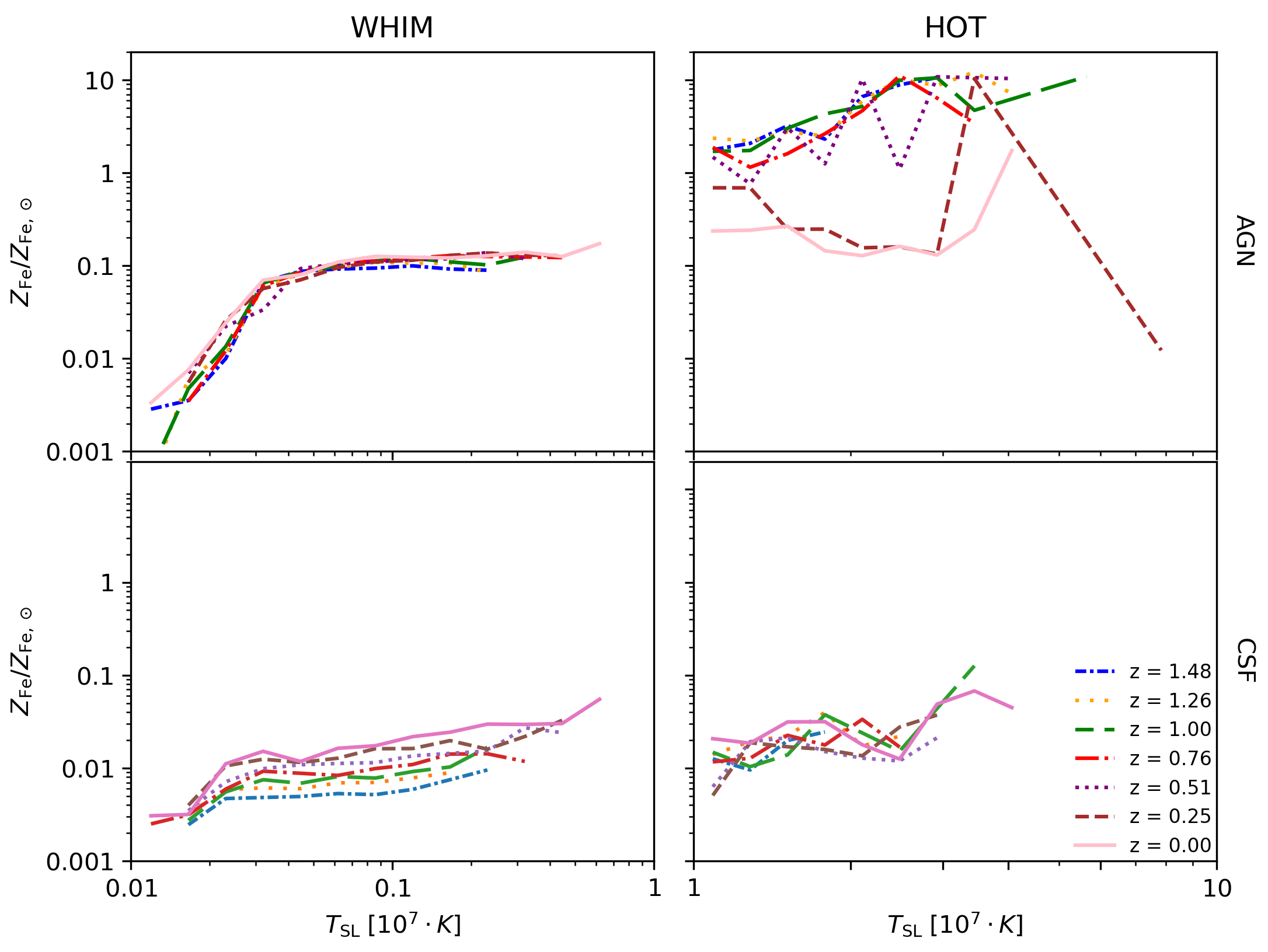}
        \caption{The median values of metallicity $Z_{Fe}$ as a function of spectroscopic-like temperature $T_{SL}$ for both WHIM and hot gas phase at seven different redshifts. The top panel is for the AGN simulations, and the bottom is for the CSF simulations. }
        \label{fig:TemperatureVsMetallicity_AGN}   
    \end{figure*}

    Focusing on the AGN model, although the hot gas inside filaments is not a relevant part of the gas mass, the filaments seem to be enriched with high metallicity hot gas. Gas in the WHIM phase is instead reaching the highest fraction of gas mass at redshift $z=0$ while having an almost constant metallicity with time. At the redshift of $\sim 1.5$, WHIM is already enriched. Its gas mass then grows by encompassing gas that gradually reaches the temperatures typical of this phase while being enriched as well. The same is not true for CSF simulations, where both the mass fraction of WHIM and its metallicity are growing in time. In this case, the WHIM phase is gradually enriched by the accumulation of gas with higher metallicities.

    To better explore the relation between the metal content and the temperature of the gas, we show in Fig. \ref{fig:TemperatureVsMetallicity_AGN} the metallicity $Z_{Fe}$ versus spectroscopic-like temperature $T_{\rm{SL}}$ for the WHIM and the hot gas phase separately. The spectroscopic-like temperature was calculated according to Eq. \ref{eq:T_SL}, where only appropriate gas particles that fall into the designated temperature and density range are included, i.e. only WHIM particles are used for the calculation of the $T_{\rm{SL}}$ for WHIM phase.  
    
    On the upper left panel of Fig. \ref{fig:TemperatureVsMetallicity_AGN} (for AGN simulations), we see that the WHIM undergoes a swift rise in metallicity from low temperatures up to approximately $T_{\rm{SL}} \approx 3\cdot 10^5 K$. The metallicity stabilizes for larger temperatures, resulting in a relatively constant metallicity. Both trends can be attributed to shock heating. According to \cite{2019MNRAS.486.3766M}, the only gas phase with metallicity lower than the WHIM is the diffuse IGM, which has a similar density to the WHIM but with temperatures below $10^5$ K. As filaments undergo shock heating, their temperature increases, causing the metal-poor diffuse IGM to transition into the WHIM phase. This transition is most evident at lower temperatures but diminishes rapidly at slightly higher temperatures. Since shock heating does not add metals to the gas, the metallicity is expected to remain constant as the temperature rises. The metallicity of the WHIM is constant across all redshifts, consistent with results in Fig. \ref{fig:Metallicity}. The hot gas phase in AGN simulations (upper right panel) displays variable median values at different redshifts. At higher redshifts, we observe a slight rise in metallicity with temperature and a gradual decline with respect to temperature at lower redshifts. Note that the majority of the filaments have a spectroscopic-like temperature below $2\cdot 10^7$ K. Similar to the discussion above, the metallicity of the hot gas is decreasing due to the shock heating. This is especially apparent at $z=0$, where we notice that the median metallicity is only slightly higher than that of WHIM and is constant at all temperatures.   
    
    On the lower panel of Fig. \ref{fig:TemperatureVsMetallicity_AGN} (CSF simulations), the metallicity of the WHIM gas phase shows a similar trend as in AGN simulations. The metallicity sharply rises until the temperature is $T_{\rm{SL}} \approx 2\cdot 10^5 K$. At higher temperatures, the metallicity rises very slowly with temperature. In time, the metallicity steadily rises at all temperatures. For the hot gas phase, the metallicity rises with higher temperatures but seemingly shows less dependence on redshift.
    
    For the WHIM phase, we found a significant diversity between the AGN and CSF simulations. In AGN simulations, the iron abundance within filaments remains unaffected by redshift, while CSF simulations show a slow rise of the values in time. This difference can be attributed to the role of AGN feedback, which was explored in the same simulations by \cite{Biffi18}, accordingly to what was discussed for Fig. \ref{fig:Metallicity} and \ref{fig:TemperatureVsMetallicity_AGN}. The AGN feedback is able to pre-enrich gas at $z>2$, and consequently, the metallicity level is already high in the AGN model for the WHIM phase.
    
\section{Conclusions}
\label{sec:conclusions}

    In this work, we studied the filaments in a set of high-resolution regions belonging to the Dianoga simulations that were centred around massive clusters. The set is composed of five different regions extracted from a parent cosmological simulation. Two of the regions were selected as "isolated", with a less massive cluster in their centre, while three of them are centred on massive clusters and densely populated with satellite clusters.
        
    We used The Sub-space Constrained Mean Shift (SCMS) algorithm \cite{Chen15} and the Sequential Chain Algorithm for Resolving Filaments (SCARF) algorithms to extract the single filaments. From there, we were able to quantify several geometrical and physical properties: length, shape, mass content for dark matter, gas and stars, mean radius, and mean density. We focused on the diffuse components (WHIM and hot gas) in the filaments, extracting their mass fractions and determining the mass-weighted metallicity and spectroscopic-like temperature.
    
    We summarize hereafter our main findings.
    \begin{enumerate}
        \item On average, the filaments inside the selected regions become longer in time. We found that longer filaments are more likely to be curved. With lower redshift, the filaments get straighter on average. 
        \item The dark matter radial profiles in filaments have different central densities for regions that evolved as isolated with respect to the densest populated ones. Moreover, we noticed that the mean radius of filaments increases during evolution, consistent with the recent results \citep{Wang24}. This behaviour is related to the fact that our regions are regions around massive galaxy clusters instead of cosmological boxes covering a fair sample of environments.
        \item We found that the mass of filaments is correlated with their length and their size. Filaments' mass increases with its length according to the power law $M \propto L^{1.7}$, similarly to what was found by \cite{Cautun14}. This indicates that the longest filaments are clearly defined and not a fortunate connection of shorter filaments. The mass also increases with a larger mean radius $\langle R \rangle$. Both correlations show a redshift dependence: at a fixed length or radius, filament masses decrease in time.
        \item We explored the evolution of the gas found in hot, WHIM and colder gas phases inside the detected filaments. The predominant gas phase in filaments is the WHIM phase. Looking at the evolution of the WHIM mass fraction, it is steadily rising towards lower redshifts. We found that the physical models do not impact the amount of baryon mass in the WHIM phase. 
        \item In general, we find that the metallicities of the WHIM and hot gas phases are, on average, higher in the AGN model than in the CSF one. 
        \item The evolution of the Iron abundance in filaments depends on the physical model used in simulations. In the case of AGN simulations, the WHIM metallicity remains constant with redshift. This is connected with the AGN feedback model, which is able to displace metals at very high redshifts \citep[e.g.][]{Biffi18}. The already enriched gas is populating the WHIM gas phase, resulting in an almost constant metallicity. The opposite is true for CSF simulations, where enriched gas is not circulated by powerful outflows, increasing Iron abundance over time.  
        \item In the case of the WHIM, metallicity is positively correlated to gas temperature, up to $T_{\rm sl} \sim 3 \cdot 10^5$ K. At higher temperatures, the metallicity remains constant.
    \end{enumerate}

    In this work, we showed that in denser regions around massive galaxy clusters, a number of filaments can be traced. The different baryon processes included in the Dianoga simulation can be used to understand the impact of AGN feedback. The trends reported in this work, especially about the gas phases and their evolution, should be investigated in more detail in future studies. To better understand the evolution of filament properties, we plan to select single filaments that persist with redshift. These filaments, possibly connected on both ends to large virialized structures, will give us a deeper insight into the regions that connect cluster outskirts and the large-scale structure. Moreover, detecting filaments near galaxy clusters in simulations will give us the opportunity to compare them with different observational data (e.g. X-ray and/or SZ observations of cluster bridges) that are currently becoming available.

\begin{acknowledgements} 

We thank the anonymous referee for the constructive comments that helped improving the paper overall, especially the presentation of the method and of the results. SI acknowledges the support from the Slovenian national research agency ARRS through grant MR-53649. SI and DF acknowledge financial support from the Slovenian Research Agency (research core funding no. P1-0188). SB is supported by: the Italian Research Center on High Performance Computing Big Data and Quantum Computing (ICSC), project funded by European Union - NextGenerationEU - and National Recovery and Resilience Plan (NRRP) - Mission 4 Component 2, within the activities of Spoke 3, Astrophysics and Cosmos Observations; by the PRIN 2022 PNRR project (202259YAF) "Space-based cosmology with Euclid: the role of High-Performance Computing". SB acknowledges partial financial support from the INFN Indark Grant. KD acknowledges support by the COMPLEX project from the European Research Council (ERC) under the European Union’s Horizon 2020 research and innovation program grant agreement ERC-2019-AdG 882679 as well as support by the Deutsche Forschungsgemeinschaft (DFG, German Research Foundation) under Germany’s Excellence Strategy - EXC-2094 - 390783311.

\end{acknowledgements}

\bibliographystyle{aa}
\bibliography{Bibliography}

\begin{appendix}

\section{The SCMS algorithm steps}
\label{app:A}

    Based on the density function $p(x)$ defined by Eq. \ref{eq:Eq1}, the SCMS algorithm shifts points to the closest density ridge, converging towards them, forming a skeleton. Here, we summarize the following steps of the SCMS algorithm \cite[see][for a detailed description]{Chen15}:
    
    \begin{enumerate}
        \item Compute the kernel density estimator $\hat{p}(x)$ via Eq. \ref{eq:Eq1}.
        \item Select a mesh $\mathcal{M}$ of points that the SCMS will shift towards density ridges. To fully cover the box volume, we used the uniformly distributed grid, with a total of $10^6$ points.
        \item Remove $x\in \mathcal{M}$ if $p(x) < \tau$, where $\tau$ is a thresholding parameter. This is called thresholding and is used to remove points in low-density regions. This significantly reduces the clutter noise and number of falsely detected filaments.
        \item For each remaining $x$, perform the SCMS:
        \begin{enumerate}
            \item Compute the Hessian matrix $H(x)$ with
                \begin{equation}
                    H(x)=\frac{1}{n}\sum_{i=1}^{n}c_{i}\bigg(\mu_{i}\mu_{i}^{T}-\frac{1}{h^{2}}\mathbb{I}\bigg),
                \end{equation}
            where 
                \begin{equation}
                    \mu_{i}=\frac{x-X_{i}}{h^{2}}, \qquad c_{i}=K\bigg(\frac{||x-X_{i}||}{h}\bigg).
                \end{equation}
            \item Perform spectral decomposition on $H(x)$. 
            \item Obtain eigenvectors corresponding to the smallest $d-1$ eigenvalues, to form $V(x)=(v_{2}(x),...,v_{d}(x))$.
            \item Update $x \longleftarrow V(x)V(x)^{T}m(x)+x$ until convergence, where 
                \begin{equation}
                    m(x) = \frac{\sum_{i=1}^{n}c_{i}X_{i}}{\sum_{i=1}^{n}c_{i}}-x
                \end{equation}
    
                is the mean shift vector.
            \end{enumerate}
        \item The final output is a collection of points that form the skeleton of the cosmic web.
    \end{enumerate}

\section{Selecting $A_0$}
\label{app:B}

In this appendix, we show the calibration of the value of the free parameter $A_0=0.5$ and, subsequently, the value of smoothing bandwidth $h$. Smoothing bandwidth $h$ controls how much we smooth the distribution of tracers when calculating $p(x)$. We are following the guidelines as written in the Appendix of \cite{Chen15}.

\begin{figure*}
    \includegraphics[width=0.95\linewidth]{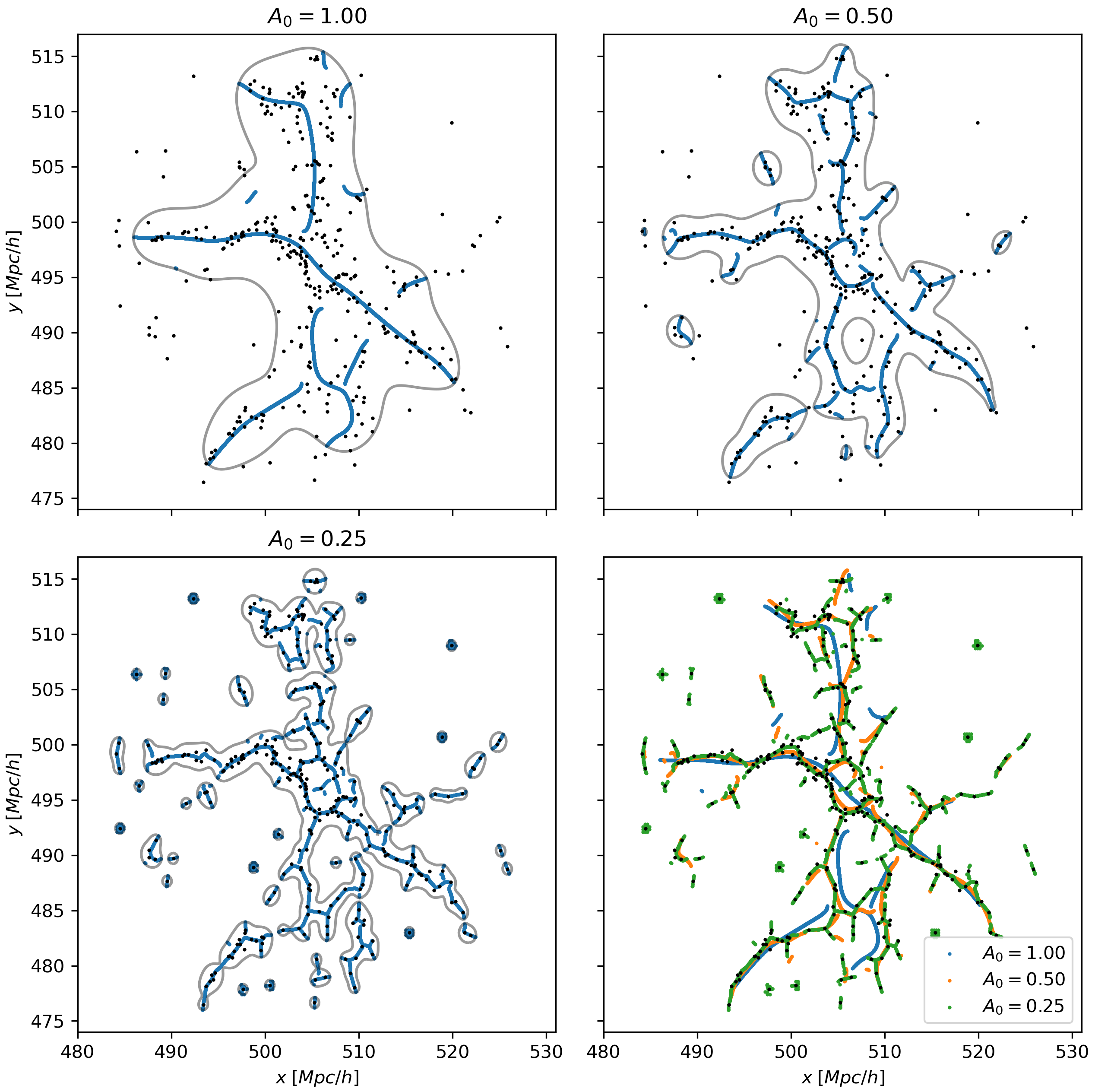}
    \caption{Skeletons, produced by SCMS, for different values of $A_0$ and therefore smoothing bandwidth $h$. The top left panel is for $A_0=1$ (corresponding to smoothing bandwidth of $2.38$ h$^{-1}$Mpc), top right panel for $A_0=0.5$ (smoothing bandwidth of $1.19$ h$^{-1}$Mpc) and bottom left panel for $A_0=0.25$ (smoothing bandwidth of $0.59$ h$^{-1}$Mpc). Tracer locations used in SCMS are marked with black dots. Contour lines in grey outline regions where $p(x)>\tau$. The bottom right panel provides a direct comparison among the three $A_0$ values, with each value distinguished by a different colour.}
    \label{fig:A_0}   
\end{figure*}

We illustrate this by using a thin slice (around $1.2$ Mpc thick) of the D6 region in AGN simulation at redshift $z=0$ (same slice as used in Fig. \ref{fig:SCMS}), centred around the largest halo in the region. The number of tracers (substructures) in this slice is $n=351$. In Fig. \ref{fig:A_0}, we show the resulting skeletons of the SCMS for different values of free parameter $A_0$. In the top left, top right and bottom left panels are the skeleton for $A_0 = 1, 0.5$ and $0.25$, respectively. These values correspond to a smoothing bandwidth of about $2.38$ h$^{-1}$ Mpc, $1.19$ h$^{-1}$Mpc and $0.59$ h$^{-1}$Mpc, respectively. The contour lines encircle the areas where $p(x) > \tau$. In the bottom right panel is the direct comparison between all three skeletons. The resulting skeletons show several distinctive differences.

Using a lower value for $A_0$ better follows the denser regions and more reliably detects filaments in lower-density areas. Still, using a value that is too low will detect more false filaments and possibly oversaturate points at the isolated tracers. We find that a good compromise is reached by adopting the value of $A_0=0.5$. Using values close to the $A_0=0.5$ does not change the skeleton significantly. 

\end{appendix}

\end{document}